# Mapping safety transitions as batteries degrade: A model-based analysis towards full-lifespan battery safety management

Xinlei Gao, Ruihe Li, Gregory J. Offer, Huizhi Wang[*]

[1] Department of Mechanical Engineering, Imperial College London, London SW7 2AZ, United Kingdom

[2] The Faraday Institution, Harwell Science and Innovation Campus, Didcot OX11 0RA, United Kingdom

[*]Corresponding author. E-mail: huizhi.wang@imperial.ac.uk

**Abstract**

Battery safety is important, yet safety limits are normally static and do not evolve as batteries degrade. Consequently, many battery systems are overengineered to meet increasingly stringent safety demands. In this work we show that failure behaviour evolves over time as batteries degrade, and discuss the challenges and opportunities to manage battery safety dynamically throughout its lifetime. We introduce the first framework for capturing how the likelihood and severity of battery failures change over time based upon the concepts of safety zones and their boundaries. Through the development of a comprehensive physics-based model that integrates multiple degradation and thermal runaway failure mechanisms, we then show how the safety zones and boundaries of a commercial 21700 battery change after varied use and how these changes may lead to false negatives with existing management strategies. Further analyses reveal that degradation mechanisms strongly affect safety characteristics, causing significant changes despite similar capacity fade, highlighting the limitations of using capacity fade alone to assess batteries' usability. By synthesising our results with literature, we map possible degradation-to-failure pathways and recommend future research needs to achieve full-lifespan battery safety management, with advanced diagnostic and modelling techniques to accurately define state-of-safety for real-world applications as key priorities.

**Introduction**

Significant advancements in energy and power densities as well as lifetime have made lithium-ion batteries a promising energy storage technology for electric vehicles and stationary storage[1–4]. Compared to other performance metrics, safety is the least addressed, yet has become a top priority due to rapidly growing battery usage in these large-scale applications[5–8]. The safety of battery systems is actively monitored and controlled by battery management systems (BMSs), which is complemented by passive safety features such as vent valves, thermal fuses, thermal barriers and reinforced casing. Current BMSs primarily rely on the State of Health (SoH) and Remaining Useful Life (RUL) based on capacity fading to assess battery reliability/usability, and apply fixed thresholds set well below "safe limits" for fault diagnosis[9–11]. Despite the seemingly conservative settings in BMSs, battery safety is inadequately addressed, occasionally leading to false negatives. As a result, many battery systems tend to be overengineered with redundant safety features and costly materials, compromising both performance and cost-effectiveness[12,13].

A root cause of the inefficiencies in active safety management is the poor understanding and definition of safe limits. There exists no unified definition of safe limits for batteries. Cai et al.[14] proposed a charging limit, set at 25% of fully lithiated graphite capacity, aimed at minimising the adverse effects of lithium (Li) plating on the maximum temperature during thermal runaway (TR). Li et al.[15] defined a zone/boundary of negative/positive capacity ratio and C-rate to restrict the maximum amount of Li deposition, thus addressing abnormal heat generation during self-heating and improving safety. Wu et al.[16] proposed a limit on heat transfer power across different states of charge (SoCs) to alleviate/delay temperature rise to reduce the pack-level TR hazard. These existing definitions focus either on specific design parameters or on operational conditions related to particular safety issues, treating their safe limits as static, unchanging values.

Although fragmented, experimental evidence in the literature has indicated that battery safety characteristics can change with degradation. Battery degradation, driven by various mechanisms (e.g., solid electrolyte interphase (SEI) layer growth, particle cracking, Li plating, electrolyte dry-out, etc.), occurs at all times, regardless of usage [17,18]. This not only results in progressive performance decay over time but also modifies material properties and structural integrity, thereby altering safety performance[19]. As illustrated in Fig. 1, excessive degradation can directly transition a battery to an unsafe state following a "DIRECT" path, which is exemplified by instances of fast-charge cycling where we showed that plated Li, a byproduct of degradation, reacts with the electrolyte, resulting in gas generation and venting failure[20]. Additionally, external abuse events (thermal, electrical, mechanical)[21] can trigger failures at any point in a battery's lifespan, with the likelihood and severity affected by usage history and aging at the time of abuse, as illustrated in the "INDIRECT" path in the figure. Feinauer et al.[22] reported a path-dependent TR behaviors, noting an increase in self-heating temperature with aging at high temperatures and a decrease at low temperatures. Through post-mortem analyses, Ren et al.[23] found that Li plating substantially reduced both self-heating and TR onset temperatures while SEI layer thickening improved TR safety performance. The dynamic changes in cell-level failure behaviors further cascade to affect the safety of battery modules and packs[24,25]. However, in contrast to the extensive studies on either battery degradation or battery safety, the change of safety performance with degradation has been ignored by most existing studies, and has never been considered in current battery management and safety countermeasures.

In this paper, we focus on the dynamic changes in safe limits throughout a battery's lifespan due to degradation, bringing evidence for the necessity of accounting for the dynamic nature of safety characteristics in battery safety management. We begin with introducing the first framework for capturing how the likelihood and severity of battery failures change over time based upon the concepts of safety zones and their boundaries, considering multiple critical events. Through the development of a comprehensive physics-based model that integrates multiple degradation and TR failure mechanisms, we then showcase how the safety zones and boundaries of a commercial 21700 battery change after being used in different ways, and how these changes may lead to inaccurate assessments of safety status with existing BMS settings. The intricate interactions among various degradation and failure mechanisms are further analysed to identify the underlying causes. By synthesising our findings with existing literature, we map possible degradation-to-failure pathways and explore technological barriers and opportunities for managing battery safety throughout its lifespan, from the first to end-of-life applications.

**Results**

*Establishing battery safety zones and their boundaries*

The framework we propose to capture the evolution of battery failure behaviours over time is illustrated in Fig. 2. To better explain how the safety zones and their boundaries are determined, the figure also plots changes in the capacity/SoH, voltage, temperature and pressure of a battery across both cycling and failure timescales. During the cycling timescale, the battery at its beginning of life (BoL) is cycled with normal voltage, temperature and pressure responses until its capacity/SoH drops below a defined failure threshold, known as the end of life (EoL). The battery cycled within the cycling timescale transitions to the failure timescale once failure is triggered either by degradation (via the "DIRECT" path) or external abuse (via the "INDERICT" path). Compared to cycling/degradation, battery failure is a rapid and aggressive process, resulting in a much shorter duration. A typical battery TR failure starts with a loss of thermal stability, leading to abnormal heat generation, which in turn heats up the battery and triggers rapid gas accumulation, causing both temperature and internal pressure to raise, as reflected by the temperature and pressure curves in the failure timescale. Eventually, venting and TR occur.

Based on the severity of events during TR failure, we categorise battery safety states into four zones: safe, transient, moderate and severe. As shown in Fig. 2, safety transition boundaries are between these zones, with the most important one being between the transient and moderate zones determined by the self-heating temperature during Accelerating Rate Calorimetry (ARC) tests[26,27], defined as "safety boundary". The "safety boundary" therefore marks the onset of abnormal heating, with the transient zone it encloses representing a margin for detecting early faults before entering the moderate zone. To prevent complete TR failure, mitigation measures are required when a battery crosses its safety boundary into the moderate zone. We also propose a "near-TR boundary" close to a threshold temperature rise (set at 1 °C min$^{-1}$ in this study) that separates the moderate and severe zones. Crossing the near-TR boundary, the rate of temperature increase surpasses the threshold temperature rise and escalates rapidly thereafter, indicating a high likelihood of complete thermal failure. The safe zone, where a battery can be safely cycled to the EoL, is bounded by the maximum operating temperature specified by its manufacturer. As batteries degrade, they may move from the safe zone into the transient zone, exhibiting instabilities in thermal and electrical responses. External abuse can push them directly into the moderate or severe zones.

To facilitate the analyses of the effects of changes of the safety zones and boundaries, two metrics, stability, and TR hazard, are used in our study to respectively evaluate the difficulty of initiating failure and the severity of failure. Changes in stability are reflected by dynamic shifts in the safety transition boundaries. When these boundaries move downward, they reduce the areas of the safety zones that remain, leading to decays in stability. Conversely, if the boundaries move upward, it indicates improvements in stability. TR hazard is measured by the maximum TR temperature shown in Fig. 2. A higher maximum TR temperature implies a higher TR hazard, and vice versa. In addition, the venting temperature ($T_v$) curve, which tracks the change in temperature at which the safety valve opens, is also included in our analyses to show the change in the difficulty of initiating venting failure throughout cycling.

***Case study analysis of evolving safety zones and boundaries and their impacts on misjudging safety status in a commercial battery***

The evolutions of the above proposed safety zones and boundaries with cycling were simulated for a commercial 21700 cylindrical cell (LG M50T). The LG M50T is a high-energy-density cell, and chosen for study because it has been well parameterised by our group and others[28–30]. This cell utilises a $SiO_x$-doped graphite anode alongside a $LiNi_{0.8}Mn_{0.1}Co_{0.1}O_2$ (NMC811) cathode with specifications and design parameters listed in Table S1, Supporting Information. It is noted that the recommended continuous charge rate for the cell is 0.3C within 0 to 25 °C and 0.7C within 25 to 45 °C. Hence, any C-rate higher than 0.3C below 25 °C and 0.7C above 25 °C is deemed as high. A standard reference performance test (RPT) was performed at every 78 cycles during cycling. Details of the model and the experiments used for model comparisons are described in the *Methods* section.

Fig. 3(a) shows how the safety zones and boundaries change over time during cycling with 0.7C charge rate and 1C discharge rate at 10 °C. The low-temperature fast-charge cycling within BMSs' limits was chosen to accelerate degradation, which can result from various sources beyond cycling in real-world applications, allowing for a clearer observation of its impact on safety characteristics. It is shown in the figure that the safety boundary interfacing between the transient and moderate zones of the battery initially drops by 10 °C at the 3rd RPT to 83.3 °C, indicating an increased likelihood of failure. Following the initial drop, the boundary stabilises until the EoL (80% SoH in this study). As for its near-TR boundary separating the moderate and severe zones, it remains at its initial value and then gradually decreases, reaching a plateau at 109.2 °C by the 3rd RPT. The shaded areas in the figure visualise the losses of stability due to degradation. Compared to 752 °C at the BoL, the maximum TR temperature decreases to 599 °C at the EoL, indicating a reduced thermal hazard of TR with degradation. The venting temperature decreases by around 3.9 °C at the EoL from the initial value of 95.7 °C, making venting failure easier to trigger. The potential consequences of ignoring these dynamic changes in the safety zones and boundaries in safety status assessments are illustrated in Fig. 3(b). In current BMSs, a static temperature threshold between 60 and 70 °C[31,32] is typically set for fault diagnosis to leave a safety margin against actual safe limits. Assuming that a threshold of $T_{fault}$ = 65 °C is applied, for this specific case, the safety margin at the BoL is 28.5 °C. However, as the battery degrades, its reduced stability and the corresponding downward shift of the safety boundary decrease the safety margin to only 18.2 °C, even at 87.5% SoH (before the EoL), indicating less room for accommodating uncertainties and errors. It is also noteworthy that the significant shrinking of the moderate zone with aging reduces the mitigation window, meaning that there is less time available to respond to warning and take actions before reaching a critical state. These should be taken into consideration when designing safety management strategies. Also shown in the figure, increased kinetics of the plated Li-electrolyte reaction (in reality can be due to an increased specific surface area of dendrite-like Li[33,34]) may further lower the safety boundary even below the fault diagnosis threshold, which possibly leads to false negatives and may explain the observed early TR initiation under 70 °C in some cycling experiments in literature[35–37].

Fig. 4 compares the dynamics of the safety zones and their boundaries under different usage paths. In addition to the 0.7C/1C cycling at 10 °C (designated as Case 1), three other usage paths, each involving a different combination of operating temperature and C-rate to trigger the dominance of specific degradation mechanisms[17], were simulated, as summarised in Table 1. It is noted in Fig. 4(a) that Cases 2, 3, and 4 exhibit similar capacity fade curves, however, they show very different changes in their safety zones and boundaries in Figs. 4(b)-(e). In Fig. 4(b), Cases 2 and 3 both show an initial increase in their safety boundaries, in contrast to the decrease observed in Case 1. In Case 2, the safety boundary gradually increases by 1.2 °C until the 6$^{th}$ RPT to 92.3 °C before dropping to 86.2 °C at the 7$^{th}$ RPT, followed by fluctuations between 86 and 91 °C until the EoL, indicating initially enhanced stability followed by a decay later in aging. In Case 3, the safety boundary increases by 5.8 °C at the 2$^{nd}$ RPT to 97 °C, and then remains constant until the EoL, indicating improved stability over cycling. Case 4 maintains the safety boundary at the initial value with a small fluctuation within 0.4 °C throughout the cycle life, suggesting minor changes in stability. In Fig. 4(c), the near-TR boundary in Case 2 follows a similar trend to Case 1, maintaining at the initial value until the 11$^{th}$ RPT and then shifting down to a plateau at 119.4 °C. Cases 3 and 4 keep their near-TR boundaries unchanged at the initial value of 145 °C. Fig. 4(d) shows that the maximum TR temperatures ($T_{max}$) in all four cases decrease, indicating reduced thermal hazards during TR. Among the four cases, Case 1 exhibits the smallest reduction in $T_{max}$, decreasing to 599 °C at the EoL. $T_{max}$ in Case 2 decreases more rapidly than Case 1, reaching a final value of 399.6 °C. Compared to Cases 1 and 2, Cases 3 and 4 show a faster and more significant decrease in $T_{max}$, reaching 335 °C and 345.2 °C at the EoL, respectively. When comparing $T_v$ in different cases in Fig. 4(e), Case 2 shows a decrease of 3.2 °C towards the EoL, which is less than Case 1; Case 3 increases $T_v$ by 2.4 °C; In Case 4, $T_v$ fluctuates within a range of 2 °C throughout the lifespan. The full diagram of the safety zones and boundaries for each of Cases 2, 3 and 4 is provided in Fig. S1 of the Supporting Information. These comparisons clearly demonstrate that the safety performance of a battery depends on how the battery has been used/degraded, suggesting that capacity fading alone is far from being sufficient for assessing battery safety status and usability.

*Probing the origins of the path dependence*

Further analyses are conducted to elucidate the processes underlying the dynamic changes in the safety zones and boundaries, as well as their path-dependence, with particular attention to the effects of four major degradation mechanisms: SEI layer growth, electrolyte dry-out, Li plating, and particle cracking. It is noted that the four degradation mechanisms are tightly coupled[17]. Particle cracking and Li plating can promote SEI layer formation and growth, and SEI layer growth consumes electrolyte solvent and can lead to electrolyte dry-out[38,39]. SEI layer growth can also cause pore blockage and thus higher electrolyte potentials to promote Li plating[40]. However, SEI layer growth leads to loss of Li inventory and increase in anode potential[41], which can impede Li plating. Figs. 5(a)-(d) compare TR heat generation variations with cycling in the four cases. The four cases exhibit different levels of decrease in total TR heat generation with cycling, explaining the observed decreases in maximum TR temperatures and thus TR hazards. To understand the reasons for the differences, the total heat generation for each case is broken down into its constituent contributions from different thermal decomposition reactions, including anode intercalated Li-electrolyte reaction (anode), plated Li-electrolyte reaction (plated Li), SEI layer decomposition (SEI), electrolyte decomposition (electrolyte), cathode decomposition (cathode), separator melting (separator) and internal short circuit (ISC)[42], as shown in Figs. 5(a)-(d). The percentage contributions of each reaction to the total heat generations for the fresh and EoL cells in the four different cases are shown in Figs. S2(a)-(e), Supporting Information. The overall decline in total TR heat generation is attributed to decreasing heat outputs from various thermal decomposition reactions over cycling, although the heat from the plated Li-electrolyte reaction can increase over time. It is shown in Figs. 5(a)-(d) that heat outputs from the anode intercalated Li-electrolyte reaction, electrolyte decomposition and internal short circuit decrease monotonically with aging, while the heats from cathode decomposition and separator melting remain unaffected. Among the four cases, Case 1 stands out with the least reduction in total heat generation, aligning with the highest TR hazard identified in Fig. 4(d). The minimum reduction is attributed to a significant increase in heat from the plated Li-electrolyte reaction, alongside a marginal decrease in heat from the anode intercalated Li-electrolyte reaction with cycling, while the heat from SEI layer decomposition fluctuates without a significant change (Fig. 5(a)). Fig. S2(b) in the Supporting Information shows that the heat from the plated Li-electrolyte reaction at the EoL accounts for 9% of the total heat generation, which is the highest among the four cases. This is ascribed to Li plating being the dominant degradation mechanism, accompanied by the lowest SEI layer growth loss and electrolyte consumption among the four cases due to low-temperature cycling, as compared in Figs. 5(e)-(j). Conversely,

Case 3 exhibits the most pronounced decline in total heat generation, explaining the lowest TR hazard in Fig. 4(d). This is attributed to a significant decrease in heat generation from the anode intercalated Li-electrolyte reaction, electrolyte decomposition and SEI layer decomposition (Fig. 5(c)), which respectively contribute 5%, 1% and 4% to the total heat generation at the EoL (Fig. S2(d), Supporting Information). Comparisons in Figs. 5(e)-(j) suggest that SEI layer growth and its associated electrolyte dry-out become dominant under the high-temperature and low C-rate conditions in Case 3. The reduced solvent concentration not only limits heat generation from the anode intercalated Li-electrolyte reaction and electrolyte decomposition, but also impedes SEI layer regeneration, thereby decreasing the heat generated from SEI layer decomposition. Case 2 shows the second lowest reduction in heat generation, along with the second-highest Li plating loss and second-lowest SEI loss among the four cases (Figs. 5(e)-(j)). In Case 2, not surprisingly, the heat from the plated Li-electrolyte reaction increases over cycling, accounting for 8% of the total heat generation as shown in Fig. S2(c) in the Supporting Information, second only to Case 1. Fig. 5(b) shows that the heat from the SEI layer decomposition in Case 2 initially increases but decreases significantly at the EoL, which can be caused by the imbalance between SEI layer decomposition and regeneration. In the middle of the cycling, the decreased solvent concentration in Case 2 is insufficient to substantially weaken SEI layer regeneration, thus leading to increased heat from SEI layer decomposition. However, SEI layer regeneration is drastically limited at the EoL, causing a significant decrease in the corresponding heat generation, similar to Case 3. In Case 4, SEI loss and electrolyte dry-out dominate over Li plating (Figs. 5(e)-(j)), resulting in heat generation patterns in Fig. 5(d) similar to Case 3, with anode intercalated Li-electrolyte reaction and electrolyte decomposition taking up 13% and 3% of the total heat generation at the EoL, respectively (Fig. S2(e), Supporting Information).

The progression of different thermal decomposition reactions with temperature, along with their associated heat and gas generation for the four different cases, is examined in Fig. 6. As shown in Fig. 6(a), the fresh cell at its initial stage of failure shows a slow progression of SEI layer decomposition and anode-electrolyte reaction under 100 °C. These reactions release some heat, nudging the cell toward its safety boundary. Subsequently, escalating exothermic reactions from the anode intercalated Li-electrolyte reaction, SEI layer decomposition, electrolyte decomposition and cathode decomposition contribute to cell heating, while the separator starts melting and absorbs some of the generated heat. With rising cell temperature, internal short circuit occurs post-separator failure, leading to a rapid release of ohmic heat. This sudden temperature surge due to the internal short circuiting intensifies other decomposition reactions until the cell reaches its maximum temperature, releasing all residual heat. Compared to the fresh cell, Figs. 6(b)-(e) show that the EoL cells in all four cases exhibit lower dimensionless concentrations for the anode intercalated Li-electrolyte reaction and electrolyte decomposition, attributable to solvent consumption during aging (Figs. 5(i) and (j)). The anode intercalated Li-electrolyte reaction in all EoL cells slows down due to the elongated diffusion pathways caused by SEI layer thickening and the reduced reaction rates resulting from electrolyte dry-out. In Case 1, due to the dominance of Li plating, the Li-electrolyte reaction initiates at ~60°C and dominates heat generation until the first exothermic peak around 100°C, as shown in Figs. 6(b1) and (b2). The reaction accounts for 52% and 80% of the total heat generated below 83 °C and 109 °C (*vs.* its safety and near-TR boundaries at 83.3 °C and 109.2 °C) (Figs. S3(b) and S4(b), Supporting Information), responsible for the downward shift in both the safety and near-TR boundaries. The rapid temperature rise below 100°C, driven by the Li-electrolyte reaction, delays the SEI layer decomposition and anode intercalated Li-electrolyte reaction (Fig. 6(b1)), resulting in a higher peak total heat generation rate at the EoL (Fig. 6(b2)). It is further supported in Fig. 6(b3) that despite the small amount of gas generation initially, the gas produced by the Li-electrolyte reaction is the main factor driving the reduction in $T_v$. Increasing the kinetics of the Li-electrolyte reaction will increase the heat generation, thus further driving down the safety and near-TR boundaries. Case 3, dominated by SEI layer growth and associated electrolyte dry-out, shows a decrease in initial concentrations for the anode intercalated Li-electrolyte reaction and electrolyte decomposition (Fig. 6(d1)). This decrease results in a lower heat generation rate during the early stage of failure in Fig. 6(d2), with only 1% of the total heat generation contributed by the anode intercalated Li-electrolyte reaction at the safety boundary (Fig. S3(d), Supporting Information), responsible for the enhanced stability and elevated safety boundary. Despite a drastic decrease in the percentage heat contribution by the anode intercalated Li-electrolyte reaction as compared in Figs. S4(a) and (d) in the Supporting Information, the near-TR boundary in Case 3 is hardly affected due to the limiting step of cathode heat generation. As a result of the high-temperature condition, the amount of plated Li in Case 3 is too small to cause any discernible difference observed in near-TR boundary and gas generation behaviours. In Case 2, the downward shift in both its safety and near-TR boundaries at the EoL can be explained by its significant Li plating loss. At the EoL, the Li-electrolyte reaction contributes 51% and 74% of the total heat generation at the safety and near-TR boundaries (Figs. S3(c) and S4(c), Supporting Information), close to those observed in Case 1. The initial upward shift in the safety boundary in Case 2 is due to the dominance of SEI layer growth and electrolyte dry-out, with the 6$^{th}$ RPT marking a transition to being dominated by Li plating, as shown in Figs. 5(e)-(j). Fig. 6(c3) shows that the continuous gas generation from the Li-electrolyte reaction reduces $T_v$ at the EoL, similar to the behaviour observed in Case 1. In Case 4, solvent consumption initially decreases heat generation (Figs. 6(e1) and (e2)). However, this decrease in heat generation is offset by the presence of plated Li, which stabilises the safety boundary and near-TR boundary at ~91 °C and ~145 °C respectively. Figs. S3(e) and S4(e) in the Supporting Information show that the heat from the Li-electrolyte reaction accounts for 43% and 34% for the EoL cell at its safety and near-TR boundaries, respectively. Li plating also offsets gas generation, resulting in a nearly constant $T_v$ in Case 4. Clearly, the path-dependent changes in safety zones and boundaries are due to their strong dependence on the relative dominance of each degradation mechanism, which differs in Cases 2, 3 and 4, even though these cases have very similar capacity fade curves.

The above findings on degradation-failure interactions, synthesised with literature, are summarised in Fig. 7, which extends our previously published degradation paths[17,28]. The figure is structured into four blocks, each representing primary and secondary degradation mechanisms, gas evolution, degradation mode, and electrochemical and safety performance. The detailed interactions between different degradation mechanisms have been discussed in our previous papers[17,28]. The interconnected degradation mechanisms pose combination effects on safety performance with significant path-dependence. Based on our findings, Li plating introduces reactive metallic Li into the closed system, which reacts with the electrolyte at temperatures below 100 °C, generating additional heat and gases during the early stages of failure. This shifts both the safety and near-TR boundaries downwards. It is worth mentioning that gas generation with Li plating mainly comprises hydrocarbon and hydrogen[43], increasing combustion and explosion risks. SEI layer growth enhances battery safety by increasing safety boundaries over time. Gases released from SEI layer decomposition are primarily carbon dioxide and carbon monoxide[44,45], with less gas hazards. Electrolyte dry-out, as a secondary degradation mechanism following SEI layer growth, also benefits safety by reducing total heat generation during failure and thus TR hazards. It consumes organic solvents, a most dangerous reactants during failure, suppressing heat generation from the anode intercalated Li-electrolyte reaction and electrolyte decomposition. According to the literature[46,47], it also allows the electrolyte to react with oxygen ($O_2$) released during cathode phase transition, converting it into less dangerous oxocarbon gases, further reducing gas hazards. The subsequent gas generation along with solvent consumption during cycling can potentially impede the ion transport and cause local Li ion concentration to increase, promoting local Li plating and thus undermining battery safety[48,49]. Cathode degradation alone has been experimentally shown in the literature[23,50] to have negligible effects on failure behaviours, and is therefore not considered in the present simulations. However, anode-cathode crosstalk can impact safety. Though not addressed in this study, it is included in Fig. 7 based on the available information from literature, due to its importance for future research. Transition metal dissolution can decrease cathode structural stability, thereby facilitating reductive gas attack and early $O_2$ release[48,50]. Hydrogen is a primary product of crosstalk, increasing combustion and explosion risks[51]. Trace water decomposition can also release hydrogen, potentially causing gas hazards. The crosstalk was found in some studies[52–55] to contribute to increased heat generation and deteriorating stability.

**Discussion**

By establishing a framework capable of capturing the evolution of battery safety characteristics over time due to degradation, and through model-based case study analyses, we highlight the importance of accounting for the evolving safety characteristics in safety management — an aspect that has been largely ignored. We demonstrate how relying on fault diagnosis/warning thresholds based on static safe limits may lead to inaccurate assessments of battery safety status and consequently suboptimal decision-making. We also reveal a strong path dependence in safety performance due to the significant influence of degradation mechanisms, where battery safety status can evolve distinctly even with similar capacity fade trajectories, making it undetectable by commonly used health indicators, SoH and RUL. This requires more reliable and effective approaches for monitoring and managing battery safety states, which calls for R&D breakthroughs in the following areas:

(1) *Diagnostic techniques*. To improve the effectiveness of safety diagnosis, it is necessary to include supplementary signals that can address the limitations in the current electrical and thermal measurements, and to develop methods for estimating safety states based on the available signals. For example, gas generation, when detected with sufficient signal resolution for fault diagnosis, whether through measuring package/electrode deformation, expansion force or internal pressure, can provide valuable insights into safety status[43,56,57]. This study indicates that even small gas generation can be useful for signalling the onset of failure process, and aiding in analysing degradation mechanisms. There is also a need to create functions and algorithms for estimating safety states that correlate with all available signal features. Increasing signal dimensionality will require the integration of additional sensors as well as more data storage and processing capabilities on board. Therefore, considerations need to be given to any trade-off involved and ensure the feasibility in real-world operating conditions.

(2) *Predictive models*. Models are central to BMSs by providing essential tools for

estimating and predicting battery states[58,59]. The model in this study represents the first published attempt to bridge degradation models with multiple coupled degradation mechanisms capable of predicting path dependent degradation with thermal runaway and failure models. However, this model cannot be used directly for real-time safety management due to several limitations: (i) it does not cover all physics, such as electrode crosstalk, limiting its accuracy and predictability to certain conditions; (ii) it involves too many parameters, requiring expensive experimental parameterisation and re-parameterisation for every different battery type; (iii) it is computationally costly due to the need for solving coupled partial differential algebraic equations. Addressing these limitations requires a deeper understanding of the intrinsic mechanisms which necessitates more fundamental experimental research. The computational limitations inherent in physics-based models call for model fusion with data-driven approaches[60–63] and advancements in high-performing and cloud computing[64–66], which hold promise for providing effective model solutions that can be constantly updated for diagnosis and prognosis purposes[67,68].

Furthermore, our study sheds light on improving battery safety for second life applications. The strong path dependence of safety characteristics necessitates a life-to-life decision-making process for any battery aged to its first EoL, where degradation mechanism/mode analyses will be conducted to aid in screening batteries for their next lives: Batteries with enhanced safety may be selected for less demanding applications in their second life due to their relatively better thermal stability, despite lower usable capacity and power; However, batteries with stability decays should be recycled or disposed. Considering the changes in safety performance when classifying second-life batteries is crucial to maximise the service life of batteries across multiple life cycles while ensuring safety. This research marks a new era in battery safety practices essential for achieving safer and more sustainable energy solutions.

**Methods**

**Degradation measurements.** To prepare aged cells for thermal abuse failure experiments and validate our degradation-safety model, degradation measurements were conducted on LG M50T cells at different ageing temperatures, covering a full SOC window from 0 to 100%. 8 fresh cells were tested, with 3 cycled at 10 °C, 3 at 40 °C, and 2 at 25 °C. At the outset, the cells were subjected to an initial break-in procedure of 5 charge-discharge cycles at 25 °C. Each cycle involved charging at 0.2C to 4.2 V, followed by a constant voltage phase until the current dropped to C/100, succeeded by discharging at 0.2C to 2.5 V. After each cycle, the cells were allowed to rest for 2 hours and 4 hours. The cells were characterised both at their BoL and after each ageing set through a RPT at 25 °C. The RPT followed two different procedures: a comprehensive procedure conducted after every even-numbered ageing set, and a concise procedure performed after each odd-numbered ageing set. Between each step of these procedures, the cells underwent a constant current-constant voltage (CC-CV) charging process at 0.3C up to 4.2 V, maintaining at 4.2 V until the current dropped to C/100, followed by a 1-hour rest period. More details about the degradation experiments refer to our previous paper[69].

**Thermal abuse failure measurements.** The cells after having reached their EoL were stored at 10 °C with 30% SOC prior to undergoing extended volume-accelerating rate calorimetry (EV-ARC) measurements at Thermal Hazard Technology (THT, UK). For comparisons, same EV-ARC measurements were conducted with fresh cells. The EV-ARC set-up used in this study is illustrated in Fig. S5, Supporting Information, which creates an adiabatic boundary around the test cell during the exothermic mode to allow the measurement of the heat generation of the cell during TR. In the EV-ARC set-up, an additional canister with a volume of 4L was placed at the centre of the EV-ARC chamber to monitor gas generation, equipped with a pressure sensor to measure gas pressure. Inside the canister, 4 K-type thermal couples were placed, with 3 attached to the cell surface at constant intervals, and the fourth placed atop the cell venting valve to measure air temperature within the canister. The temperature and pressure signals were recorded at a frequency of 1 Hz. All test cells were charged to 100% SOC using the CC-CV protocol before the EV-ARC tests. The TR behaviours were tested using the Heat-Wait-Seek mode, with testing parameters listed in Table S2, Supporting Information.

**Physics-based degradation-safety model.** To systematically study the degradation effects on battery safety performance, a physics-based degradation-safety model was developed by coupling a degradation sub-model we had previously developed[28,39] with a TR failure sub-model. The degradation sub-model is based on the classical Doyle-Fuller-Newman (DFN) model[70] and includes strongly coupled models of SEI layer growth, electrolyte dry-out, Li plating, particle cracking and associated loss of active materials. The DFN model solves charge and mass conservation in the electrodes and electrolyte of a battery cell, with the total current density contributed by both Li (de)intercalation and degradation side reactions (details see Supplementary Method, Supporting Information). The SEI layer growth process is modelled using the diffusion limited SEI model, which assumes that the thickness growth rate of SEI layer is limited by the solvent transport through the outer SEI layer[71]:

$$\frac{\partial L_{\text{SEI}}}{\partial t} = -\frac{1}{2} N_{\text{sol}} \bar{v}_{\text{SEI}} \qquad (1)$$

with the solvent flux, $N_{sol}$, described by

$$N_{sol} = -D_{sol}(T)\frac{\partial c_{sol}}{\partial l} \tag{2}$$

where $\bar{v}_{SEI}$ is SEI partial molar volume, $c_{sol}$ is the solvent concentration, $D_{sol}(T)$ is the temperature-dependent diffusivity, and $l$ represents the location along the SEI layer.

The voltage drop associated with the SEI layer growth, $\eta_{SEI}$, is determined by:

$$\eta_{SEI} = \rho_{SEI}L_{SEI}\frac{J_{tot}}{a_n} \tag{3}$$

where $a_n$ is specific surface area, $\rho_{SEI}$ is ohmic resistivity and $J_{tot}$ is total current density.

The solvent consumption resulting from the SEI layer growth is modelled following the approach in our previous work[39], wherein the amount of solvent consumed over a time interval $\Delta t$ is calculated and incorporated into the DFN model. By assuming that the electrolyte reservoir is empty, the change of cell cross-sectional area $A_{cell}$ due to solvent consumption is calculated using Eqs. (4)-(7):

$$\Delta n_{SEI} = \frac{A_{cell}(t)}{\bar{v}_{SEI}}\int_0^{L_n} a_n[L_{SEI}(t+\Delta t) - L_{SEI}]dx \tag{4}$$

$$Ratio_{dry} = \frac{V_e(t) + \Delta n_{SEI}(\bar{v}_{SEI} - 2\bar{v}_{EC})}{V_e(t)} \tag{5}$$

$$V_e(t) = A_{cell}(t)[L_n\varepsilon_n(t) + L_{sep}\varepsilon_{sep} + L_p\varepsilon_p] \tag{6}$$

$$A_{cell}(t+\Delta t) = Ratio_{dry}A_{cell}(t) \tag{7}$$

where $\Delta n_{SEI}$ is the amount change of SEI, $\bar{v}_{EC}$ is the partial molar volume for the solvent of ethylene carbonate, and $Ratio_{dry}$ denotes the shrinking ratio of the cross-sectional area.

Li formation, stripping and the subsequent decay of plated Li into dead Li are simulated using a partially reversible Li plating model, where the concentration change of plated Li is correlated with the Li plating current density, $J_{Li}$, and the rate of conversion of plated Li to "dead Li", $\gamma_{Li} c_{Li}$, by[28]

$$\frac{\partial c_{Li}}{\partial t} = \frac{-J_{Li}}{F} - \gamma_{Li} c_{Li} \tag{8}$$

with the Li plating current density determined from the Butler-Volmer equation, and the decay rate to "dead Li", $\gamma_{Li}$, assumed to be diffusion limited and dependent on the SEI thickness by:

$$\gamma = \gamma_0 \frac{L_{SEI,0}}{L_{SEI}} \tag{9}$$

The porosity change due to SEI layer growth and partially reversible Li plating is determined by

$$\frac{d\varepsilon}{dt} = -a_n \cdot \frac{dL_{tot}}{dt} \tag{10}$$

where $L_{tot}$ is the total thickness of deposition calculated as

$$L_{tot} = L_{SEI} + c_{Li} \frac{\bar{v}_{Li}}{a_n} + c_{dl} \frac{\bar{v}_{Li}}{a_n} \tag{11}$$

where $c_{Li}$ and $c_{dl}$ denote the concentrations of active plated Li and dead Li, respectively.

Electrode particle fracture and crack propagation are unified into a single stress-based model[72], with the radial stress $\sigma_r$ and tangential stress $\sigma_t$ solved by

$$\sigma_r = \frac{2\Omega\psi}{(1-\nu)} [c_{avg}(R_i) - c_{avg}(r)] \tag{12}$$

$$\sigma_t = \frac{\Omega\psi}{(1-v)}[2c_{\text{avg}}(R_i) + c_{\text{avg}}(r) - \bar{c}/3)] \qquad (13)$$

where $\Omega$ denotes the partial molar volume, $\psi$ is the Young's modulus, $v$ is the Possion's ratio, $R$ denotes the radius of the particle, $c_{\text{avg}}(r)$ is the average Li ion conceneration and $\bar{c}$ is the difference between the Li concentration and reference value.

The fatigue crack growth is solved by Paris' law as below

$$\frac{\text{d}l_{\text{cr}}}{\text{d}N} = \frac{k_{\text{cr}}}{t_0}\left(\sigma_t b_{\text{cr}}\sqrt{\pi l_{\text{cr}}}\right)^{\lambda_{\text{cr}}}, \text{ when } \sigma_t > 0 \qquad (14)$$

where $t_0$ is the cycle number, $k_{\text{cr}}$ is the correction factor for stress intensity, $b_{\text{cr}}$ and $\lambda_{\text{cr}}$ are constants.

The loss of active material (LAM) due to stress is calculated by[28]

$$\frac{\partial \varepsilon_{s,k}}{\partial t} = \frac{\beta}{t_0}\left(\frac{\sigma_h}{\sigma_c}\right)^{\lambda_{\text{LAM}}}, \text{ when } \sigma_h > 0 \qquad (15)$$

$$\sigma_h = (\sigma_r + 2\sigma_t)/3 \qquad (16)$$

where $\varepsilon_{a,k}$ is the volume fraction for active materials, $\beta$ and $\lambda_{\text{LAM}}$ are constants of LAM proportional and exponential terms, and $\sigma_c$ is the critical stress.

The TR failure sub-model considers thermal decomposition reactions and their associated heat and gas generation, pressure buildup and vent flow dynamics. The energy conservation equation during TR is as follows[73]:

$$m_{\text{cell}}Cp_{\text{cell}}\frac{dT_{\text{cell}}}{dt} = \sum \dot{Q}_i + \dot{Q}_{\text{diss}} + \dot{Q}_{\text{short}} + \dot{Q}_{\text{evap}} + \dot{m}_{\text{vent}}\Delta H_{\text{vent}} \qquad (17)$$

where $\dot{Q}_i$, $\dot{Q}_{\text{diss}}$, $\dot{Q}_{\text{short}}$, $\dot{Q}_{\text{evap}}$, and $\dot{m}_{\text{vent}} h_{\text{vent}}$ represent heat generation rate from the reaction $i$, heat dissipation rate, heat generation from internal short circuit, evaporation heat from solvents and energy loss to ejecta, respectively. The reaction heat generation rate is determined by

$$\dot{Q}_i(t) = m_i \Delta H_i \frac{dc_i^*(t)}{dt} \tag{18}$$

with the reaction rate calculated by the Arrhenius law:

$$\frac{dc_i^*(t)}{dt} = B_i \exp\left(-\frac{E_i}{RT_{\text{cell}}(t)}\right) f_i\left(c_i^*(t)\right) \tag{19}$$

where $R$ denotes the gas constant, $B_i$, $c_i^*(t)$ and $E_i$ denote the frequency, normalised reactant concentration and activation energy of the reaction $i$, which connect to the degradation sub-model. $f_i\left(c_i^*(t)\right)$ is a modification factor, varying for different reactions and linking back to the degradation sub-model.

The calculation of the internal pressure within the battery prior to venting follows Dalton's law, accounting for contributions from both generated gases and electrolyte vapour:

$$P_{\text{tot}} = P_{\text{gas}} + P_{\text{elec}} \tag{20}$$

with the partial pressures of generated gases, $P_{\text{gas}}$, derived from the ideal gas law alongside kinetics of gas-generating reactions, and the electrolyte vapor pressure determined as the saturated pressure using Antoine's equation[74]. At each time step, the total internal pressure is compared to the vent pressure (2 MPa for LGM50T cells). If it is lower than the vent pressure, the partial pressures calculations are repeated. If the pressure exceeds the vent pressure, the vent valve opens, and the dynamics of vent flow are calculated as an isentropic nozzle flow, considering two flow conditions: subsonic flow and chocked flow as determined by the inequality[75],

$$\frac{P_{\text{amb}}}{P_{\text{cell}}} > \left(\frac{2}{\theta+1}\right)^{\frac{\theta}{\theta-1}} \tag{21}$$

where $\theta$ is the heat capacity ratio of venting gases. The flow dynamics calculations yield the vent gas velocity, facilitating the calculation of the vent mass flow rate. The mass flow rate is subsequently substituted back into Eq. (17) to calculate the heat carried by the vent mass and thus update the cell temperature. The degradation sub-model is implemented in the open-source package Python Battery Mathematical Modelling (PyBaMM)[76], and the TR failure sub-model is coded in Python and coupled with the PyBaMM degradation model. A full list of the model equations, boundary conditions and input parameters are provided in Supplementary Method, Supporting Information. The model was validated against the experimental results prior to being used for case study analyses. Our simulation results show a good agreement with the measurement results as shown in Figs. S8 and S9, Supporting Information.

## Author contributions

**Xinlei Gao:** Conceptualisation, methodology, data curation, investigation, writing - original draft & editing, review & editing; **Ruihe Li:** Methodology, data curation; **Gregory J. Offer:** Conceptualisation, supervision, funding acquisition, review & editing; **Huizhi Wang:** Conceptualisation, supervision, investigation, funding acquisition, review & editing.

## Declaration of competing interest

The authors declare no competing interests.

## Acknowledgements

The authors would like to thank Thermal Hazard Technology (THT) for their assistance in battery abuse testing. This work is supported by the UK Faraday Institution BESAFE project (FIRG038) and Multi-Scale Modelling project (FIRG025) and the Imperial College President's PhD Scholarship Scheme (funded by EPSRC). For the purpose of open access, the authors have applied a 'Creative Commons Attribution (CC BY) licence' (where permitted by UKRI, 'Open Government Licence' or 'Creative Commons Attribution No-derivatives (CC BY-ND) licence' may be stated instead) to any Author Accepted Manuscript version arising.

**References**

1. Grey, C. P. & Hall, D. S. Prospects for lithium-ion batteries and beyond—a 2030 vision. *Nat Commun* **11**, 1–4 (2020).

2. Davies, D. M. *et al.* Combined economic and technological evaluation of battery energy storage for grid applications. *Nat Energy* **4**, 42–50 (2018).

3. Xiao, J., Shi, F., Glossmann, T., Burnett, C. & Liu, Z. From laboratory innovations to materials manufacturing for lithium-based batteries. *Nat Energy* **8**, 329–339 (2023).

4. Aguilar Lopez, F., Lauinger, D., Vuille, F. & Müller, D. B. On the potential of vehicle-to-grid and second-life batteries to provide energy and material security. *Nat Commun* **15**, 1–11 (2024).

5. Deng, J., Bae, C., Marcicki, J., Masias, A. & Miller, T. Safety modelling and testing of lithium-ion batteries in electrified vehicles. *Nat Energy* **3**, 261–266 (2018).

6. Finegan, D. P. & Cooper, S. J. Battery Safety: Data-Driven Prediction of Failure.

*Joule* **3**, 2599–2601 (2019).

7.  Wang, J. *et al.* Rapid temperature-responsive thermal regulator for safety management of battery modules. *Nat Energy* **9**, 939–946 (2024).

8.  Huang, L. *et al.* Thermal runaway routes of large-format lithium-sulfur pouch cell batteries. *Joule* **6**, 906–922 (2022).

9.  Che, Y., Hu, X., Lin, X., Guo, J. & Teodorescu, R. Health prognostics for lithium-ion batteries: mechanisms, methods, and prospects. *Energy Environ Sci* **16**, 338–371 (2023).

10. Hu, X., Xu, L., Lin, X. & Pecht, M. Battery Lifetime Prognostics. *Joule* **4**, 310–346 (2020).

11. Li, W., Zhu, J., Xia, Y., Gorji, M. B. & Wierzbicki, T. Data-Driven Safety Envelope of Lithium-Ion Batteries for Electric Vehicles. *Joule* **3**, 2703–2715 (2019).

12. Campbell, I. D. *et al.* Optimising lithium-ion cell design for plug-in hybrid and battery electric vehicles. *J Energy Storage* **22**, 228–238 (2019).

13. Li, S., Zhang, C., Zhao, Y., Offer, G. J. & Marinescu, M. Effect of thermal gradients on inhomogeneous degradation in lithium-ion batteries. *Communications Engineering* **2**, 74 (2023).

14. Cai, W. *et al.* The Boundary of Lithium Plating in Graphite Electrode for Safe Lithium-Ion Batteries. *Angewandte Chemie - International Edition* **60**, 13007–13012 (2021).

15. Li, H., Ji, W., Zhang, P. & Zhao, J. Safety boundary of power battery based on quantitative lithium deposition. *J Energy Storage* **52**, 104789 (2022).

16. Wu, H. *et al.* Thermal safety boundary of lithium-ion battery at different state of charge. *Journal of Energy Chemistry* **91**, 59–72 (2024).

17. Edge, J. S. *et al.* Lithium ion battery degradation: what you need to know. *Physical Chemistry Chemical Physics* **23**, 8200–8221 (2021).

18. Birkl, C. R., Roberts, M. R., McTurk, E., Bruce, P. G. & Howey, D. A. Degradation diagnostics for lithium ion cells. *J Power Sources* **341**, 373–386 (2017).

19. Finegan, D. P. *et al.* The battery failure databank: Insights from an open-access database of thermal runaway behaviors of Li-ion cells and a resource for benchmarking risks. *J Power Sources* **597**, 234106 (2024).

20. Li, Y. *et al.* Direct venting during fast charging of lithium-ion batteries. *J Power Sources* **592**, 233926 (2024).

21. Feng, X., Ren, D., He, X. & Ouyang, M. Mitigating Thermal Runaway of Lithium-Ion Batteries. *Joule* **4**, 743–770 (2020).

22. Feinauer, M. *et al.* Change of safety by main aging mechanism – A multi-sensor accelerating rate calorimetry study with commercial Li-ion pouch cells. *J Power Sources* **570**, 233046 (2023).

23. Ren, D. *et al.* A comparative investigation of aging effects on thermal runaway behavior of lithium-ion batteries. *eTransportation* **2**, 100034 (2019).

24. Radhakrishnan, A. N. P. *et al.* Quantitative spatiotemporal mapping of thermal runaway propagation rates in lithium-ion cells using cross-correlated Gabor filtering. *Energy Environ Sci* **15**, 3503–3518 (2022).

25. Huang, W., Feng, X., Han, X., Zhang, W. & Jiang, F. Questions and Answers Relating to Lithium-Ion Battery Safety Issues. *Cell Rep Phys Sci* **2**, 100285 (2021).

26. Börger, A., Mertens, J. & Wenzl, H. Thermal runaway and thermal runaway propagation in batteries: What do we talk about? *J Energy Storage* **24**, 100649 (2019).

27. Feng, X. *et al.* Thermal runaway mechanism of lithium ion battery for electric vehicles: A review. *Energy Storage Mater.* **10**, 246 (2018).

28. O'Kane, S. E. J. *et al.* Lithium-ion battery degradation: how to model it. *Physical Chemistry Chemical Physics* **24**, 7909–7922 (2022).

29. O'Regan, K., Brosa Planella, F., Widanage, W. D. & Kendrick, E. Thermal-electrochemical parameters of a high energy lithium-ion cylindrical battery. *Electrochim Acta* **425**, 140700 (2022).

30. Chen, C.-H. *et al.* Development of Experimental Techniques for Parameterization of Multi-scale Lithium-ion Battery Models. *J Electrochem Soc* **167**, 080534 (2020).

31. Hu, X. *et al.* Advanced Fault Diagnosis for Lithium-Ion Battery Systems: A Review of Fault Mechanisms, Fault Features, and Diagnosis Procedures. *IEEE*

*Industrial Electronics Magazine* **14**, 65–91 (2020).

32. Xiong, R., Sun, W., Yu, Q. & Sun, F. Research progress, challenges and prospects of fault diagnosis on battery system of electric vehicles. *Appl Energy* **279**, 115855 (2020).

33. Mao, S. *et al.* In situ evaluation and manipulation of lithium plating morphology enabling safe and long-life lithium-ion batteries. *InfoMat* (2024) doi:https://doi.org/10.1002/inf2.12612.

34. Gao, X. *et al.* Thermodynamic Understanding of Li-Dendrite Formation. *Joule* **4**, 1864–1879 (2020).

35. Li, Y. *et al.* Thermal Runaway Triggered by Plated Lithium on the Anode after Fast Charging. *ACS Appl Mater Interfaces* **11**, 46839–46850 (2019).

36. Li, Y. *et al.* Battery eruption triggered by plated lithium on an anode during thermal runaway after fast charging. *Energy* **239**, 122097 (2022).

37. Zhang, G. *et al.* Revealing the Impact of Fast Charge Cycling on the Thermal Safety of Lithium-Ion Batteries. *ACS Appl Energy Mater* **5**, 7056–7068 (2022).

38. Gao, T. *et al.* Interplay of lithium intercalation and plating on a single graphite particle. *Joule* **5**, 393–414 (2021).

39. Li, R., O'Kane, S., Marinescu, M. & Offer, G. J. Modelling Solvent Consumption from SEI Layer Growth in Lithium-Ion Batteries. *J Electrochem Soc* **169**, 060516 (2022).

40. Liu, Q. *et al.* Understanding undesirable anode lithium plating issues in lithium-

ion batteries. *RSC Adv* **6**, 88683–88700 (2016).

41. Raj, T., Wang, A. A., Monroe, C. W. & Howey, D. A. Investigation of Path-Dependent Degradation in Lithium-Ion Batteries**. *Batter Supercaps* **3**, 1377–1385 (2020).

42. Feng, X., Lu, L., Ouyang, M., Li, J. & He, X. A 3D thermal runaway propagation model for a large format lithium ion battery module. *Energy* **115**, 194–208 (2016).

43. Jin, Y. *et al.* Detection of Micro-Scale Li Dendrite via H2 Gas Capture for Early Safety Warning. *Joule* **4**, 1714–1729 (2020).

44. Bernhard, R., Metzger, M. & Gasteiger, H. A. Gas Evolution at Graphite Anodes Depending on Electrolyte Water Content and SEI Quality Studied by On-Line Electrochemical Mass Spectrometry. *J Electrochem Soc* **162**, A1984–A1989 (2015).

45. Stock, S. *et al.* Operando Analysis of the Gassing and Swelling Behavior of Lithium-ion Pouch Cells during Formation. *J Electrochem Soc* **170**, 060539 (2023).

46. Wei, G. *et al.* A comprehensive insight into the thermal runaway issues in the view of lithium-ion battery intrinsic safety performance and venting gas explosion hazards. *Appl Energy* **349**, 121651 (2023).

47. Liu, P. *et al.* Revealing Lithium Battery Gas Generation for Safer Practical Applications. *Adv Funct Mater* **32**, 2208586 (2022).

48. Zhang, G., Shen, W. & Wei, X. Lithium-ion battery thermal safety evolution

during high-temperature nonlinear aging. *Fuel* **362**, 130845 (2024).

49. Zhang, G. *et al.* Thermal characteristic evolution of lithium−ion batteries during the whole lifecycle. *Journal of Energy Chemistry* **92**, 534–547 (2024).

50. Zhang, G. *et al.* Research on the impact of high-temperature aging on the thermal safety of lithium-ion batteries. *Journal of Energy Chemistry* **87**, 378–389 (2023).

51. Wang, X. *et al.* Ni crossover catalysis: truth of hydrogen evolution in Ni-rich cathode-based lithium-ion batteries. *Energy Environ Sci* **16**, 1200–1209 (2023).

52. Song, Y. *et al.* The significance of mitigating crosstalk in lithium-ion batteries: a review. *Energy Environ Sci* **16**, 1943–1963 (2023).

53. Meunier, V., Leal De Souza, M., Morcrette, M. & Grimaud, A. Design of workflows for crosstalk detection and lifetime deviation onset in Li-ion batteries. *Joule* **7**, 42–56 (2023).

54. Wang, Y. *et al.* Reductive gas manipulation at early self-heating stage enables controllable battery thermal failure. *Joule* **6**, 2810–2820 (2022).

55. Zhou, H. *et al.* Effect of electrode crosstalk on heat release in lithium-ion batteries under thermal abuse scenarios. *Energy Storage Mater* **44**, 326–341 (2022).

56. Bezsonov, I. I., Waller, G. H., Ko, J. & Nadimpalli, S. P. V. In operando measurement of surface strain of 18650 Li-ion cells during cycling. *J Power Sources* **592**, 233915 (2024).

57. Willenberg, L. K., Dechent, P., Fuchs, G., Sauer, D. U. & Figgemeier, E. High-

Precision Monitoring of Volume Change of Commercial Lithium-Ion Batteries by Using Strain Gauges. *Sustainability* **12**, 557 (2020).

58. Dubarry, M., Truchot, C. & Liaw, B. Y. Synthesize battery degradation modes via a diagnostic and prognostic model. *J Power Sources* **219**, 204–216 (2012).

59. Sulzer, V. *et al.* The challenge and opportunity of battery lifetime prediction from field data. *Joule* **5**, 1934–1955 (2021).

60. Wang, Y. *et al.* Temperature excavation to boost machine learning battery thermochemical predictions. *Joule* (2024) doi:https://doi.org/10.1016/j.joule.2024.07.002.

61. Ma, G. *et al.* Real-time personalized health status prediction of lithium-ion batteries using deep transfer learning. *Energy Environ Sci* **15**, 4083–4094 (2022).

62. Finegan, D. P. *et al.* The Application of Data-Driven Methods and Physics-Based Learning for Improving Battery Safety. *Joule* **5**, 316–329 (2020).

63. Chen, B. R. *et al.* Battery aging mode identification across NMC compositions and designs using machine learning. *Joule* **6**, 2776–2793 (2022).

64. Yang, S. *et al.* CHAIN: Cyber Hierarchy and Interactional Network Enabling Digital Solution for Battery Full-Lifespan Management. *Matter* **3**, 1–15 (2020).

65. Gao, X. *et al.* Designed high-performance lithium-ion battery electrodes using a novel hybrid model-data driven approach. *Energy Storage Mater* **36**, 435–458 (2021).

66. Zhang, J. *et al.* Realistic fault detection of li-ion battery via dynamical deep

learning. *Nat Commun* **14**, 5940 (2023).

67. Tao, F. & Qi, Q. Make more digital twins. *Nature* **573**, 490–491 (2019).

68. Dubarry, M., Howey, D. & Wu, B. Enabling battery digital twins at the industrial scale. *Joule* **7**, 1134–1144 (2023).

69. Kirkaldy, N., Samieian, M. A., Offer, G. J., Marinescu, M. & Patel, Y. Lithium-Ion Battery Degradation: Measuring Rapid Loss of Active Silicon in Silicon–Graphite Composite Electrodes. *ACS Appl Energy Mater* **5**, 13367–13376 (2022).

70. Doyle, M., Fuller, T. F. & Newman, J. Modeling of Galvanostatic Charge and Discharge of the Lithium/Polymer/Insertion Cell. *J Electrochem Soc* **140**, 1526–1533 (1993).

71. Single, F., Latz, A. & Horstmann, B. Identifying the Mechanism of Continued Growth of the Solid–Electrolyte Interphase. *ChemSusChem* **11**, 1950–1955 (2018).

72. Zhang, X., Shyy, W. & Sastry, A. M. Numerical Simulation of Intercalation-Induced Stress in Li-Ion Battery Electrode Particles. *J Electrochem Soc* **154**, A910 (2007).

73. He, C. X., Yue, Q. L., Chen, Q. & Zhao, T. S. Modeling thermal runaway of lithium-ion batteries with a venting process. *Appl Energy* **327**, 120110 (2022).

74. Zhang, X., Zuo, J. & Jian, C. Experimental isobaric vapor-liquid equilibrium for binary systems of ethyl methyl carbonate + methanol, + ethanol, + dimethyl carbonate, or + diethyl carbonate at 101.3 kPa. *J Chem Eng Data* **55**, 4896–4902
34

(2010).

75. Kong, D., Wang, G., Ping, P. & Wen, J. A coupled conjugate heat transfer and CFD model for the thermal runaway evolution and jet fire of 18650 lithium-ion battery under thermal abuse. *eTransportation* **12**, 100157 (2022).

76. Sulzer, V., Marquis, S. G., Timms, R., Robinson, M. & Chapman, S. J. Python Battery Mathematical Modelling (PyBaMM). *J Open Res Softw* **9**, 14 (2021).

**List of tables**

**Table 1.** Parameter settings for four different degradation paths with different combinations of operating temperature and C-rate to trigger the dominance of specific degradation mechanisms. The arrows indicate values outside the manufacturer's recommended range as in the cell specification sheet: "↑" means the value is higher than recommended, and "↓" means it is lower.

| Case | $T$ (°C) | Charge rate | Discharge rate | Cycle life (cycle) | Maximum RPT | Expected dominant degradation mechanism |
|---|---|---|---|---|---|---|
| 1 | 10 | 0.7C ↑ | 1C | 390 | 6 | Li plating |
| 2 | 25 | 1C ↑ | 1C | 936 | 13 | Li plating; SEI growth; Electrolyte dry-out |
| 3 | 40 | 0.3C ↓ | 1C | 1170 | 16 | SEI growth; Electrolyte dry-out |
| 4 | 40 | 1.5C ↑ | 1C | 1170 | 16 | Li plating; SEI growth; Electrolyte |

**List of figures**

**Fig. 1** The dynamic evolution of battery safety with degradation. A battery degrades from its BoL to EoL with various degradation mechanisms accumulated inside. Two paths, namely direct and indirect paths, lead to abnormal heat and gas generation, transitioning the battery from a safe to an unsafe state. The cell-level failures can cascade to affect module/pack-level safety.

**Fig. 2** Proposed four zones for battery safety states with dynamic transition boundaries between them, considering multiple critical events during failure.

**Fig. 3** (a) Dynamic evolutions of the safety zones and boundaries in a commercial battery, and (b) their impacts on misjudging safety status of the battery. The battery was cycled at a charge rate of 0.7C and a discharge rate of 1C at 10 ºC.

**Fig. 4** Changes in (a) capacity, (b) the safety boundary and the transient zone, (c) the near-TR boundary and the moderate zone, (d) the maximum TR temperature and the severe zone, and (e) venting temperature of a battery after being used in four different ways.

**Fig. 5** Total TR heat generation and its constituent contributions from different thermal decomposition reactions with cycling for (a) Case 1, (b) Case 2, (c) Case 3 and (d) Case 4, alongside the (e) SEI growth-induced capacity loss, (f) SEI layer thickness change, (g) Li plating-induced capacity loss, (h) plated dead Li amount, (i) solvent loss, and (j) solvent concentration change throughout cycling in the four cases.

**Fig. 6** Comparative analyses of thermal decomposition reaction kinetics between the (a) fresh cell and EoL cells in (b) Case 1, (c) Case 2, (d) Case 3 and (e) Case 4. Subfigures 1, 2 and 3 present dimensionless concentrations, heat generation powers and amounts of generated gases from different thermal decomposition reactions. Dashed lines indicate the locations of the safety and near-TR boundaries.

**Fig. 7** Mapping degradation-safety pathways.

Fig. 1

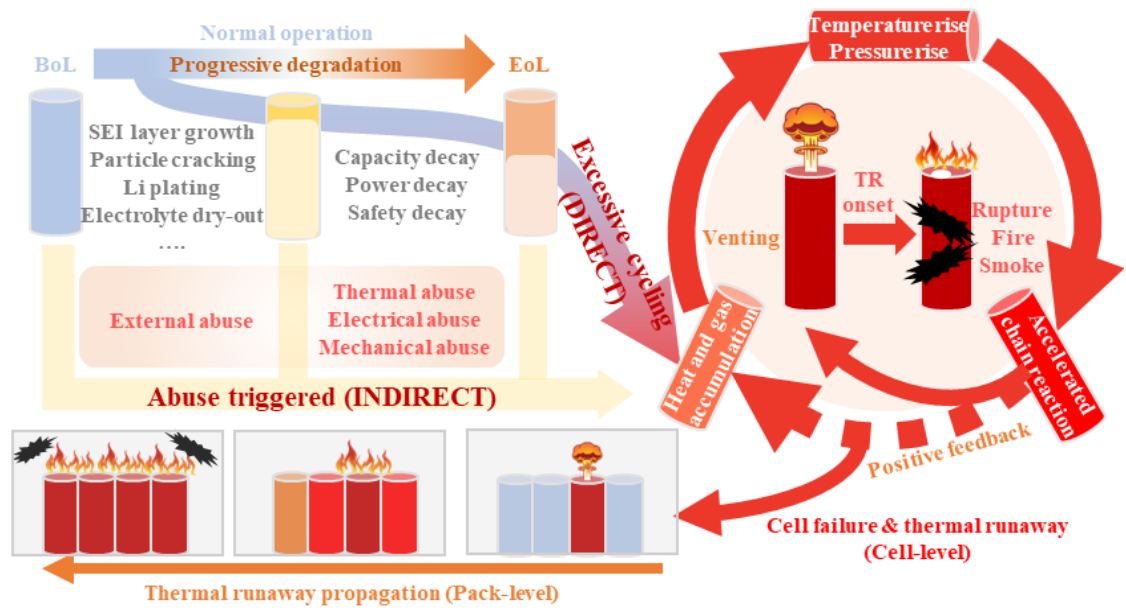

Fig. 2

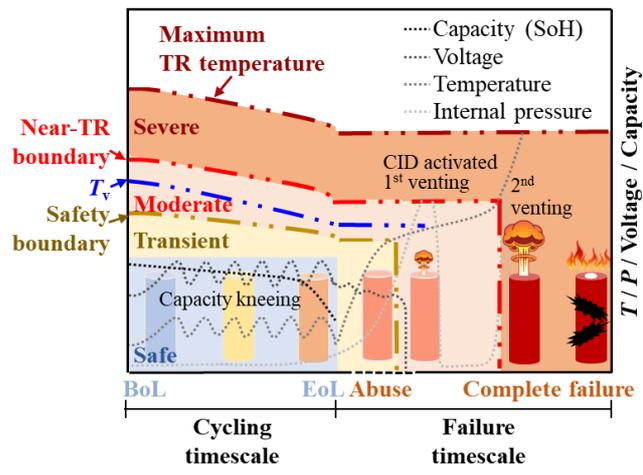

Fig. 3

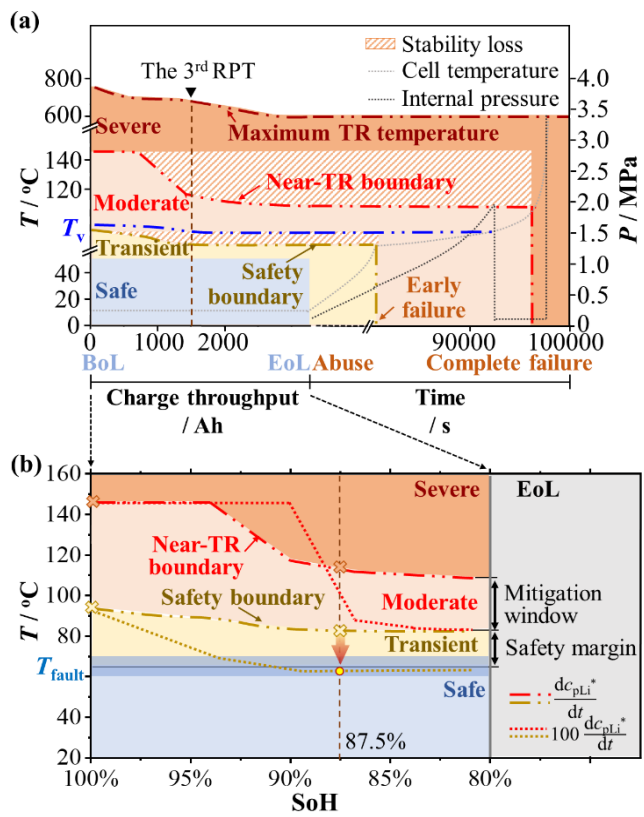

Fig. 4

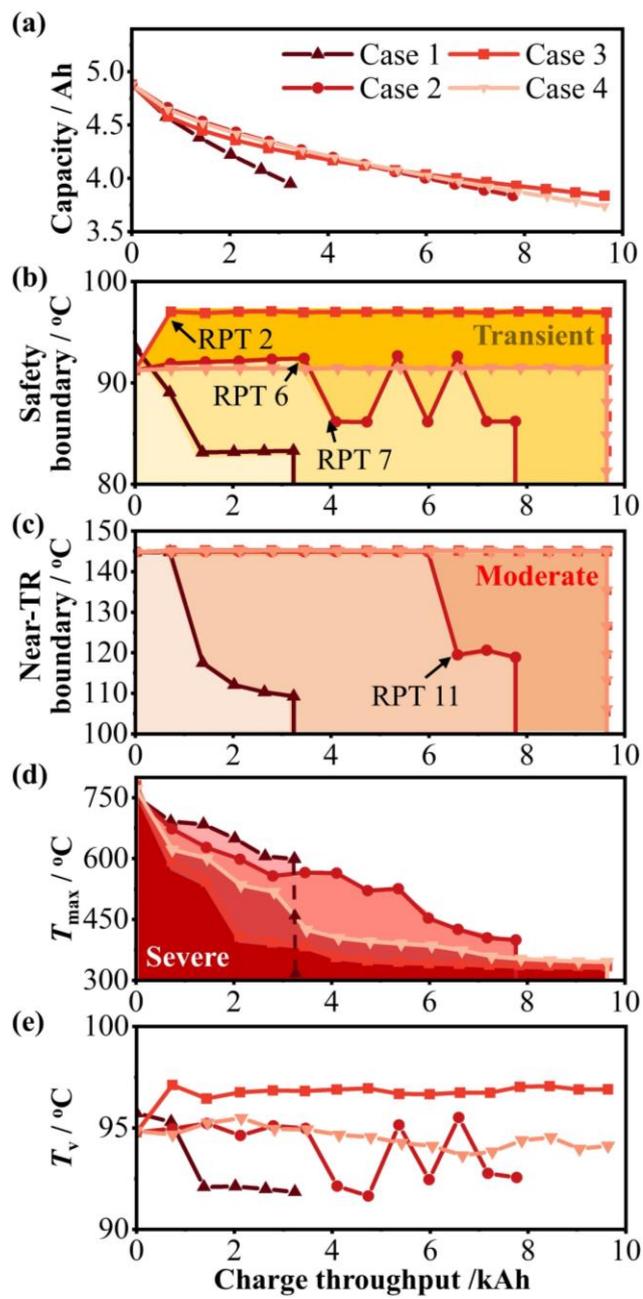

Fig. 5

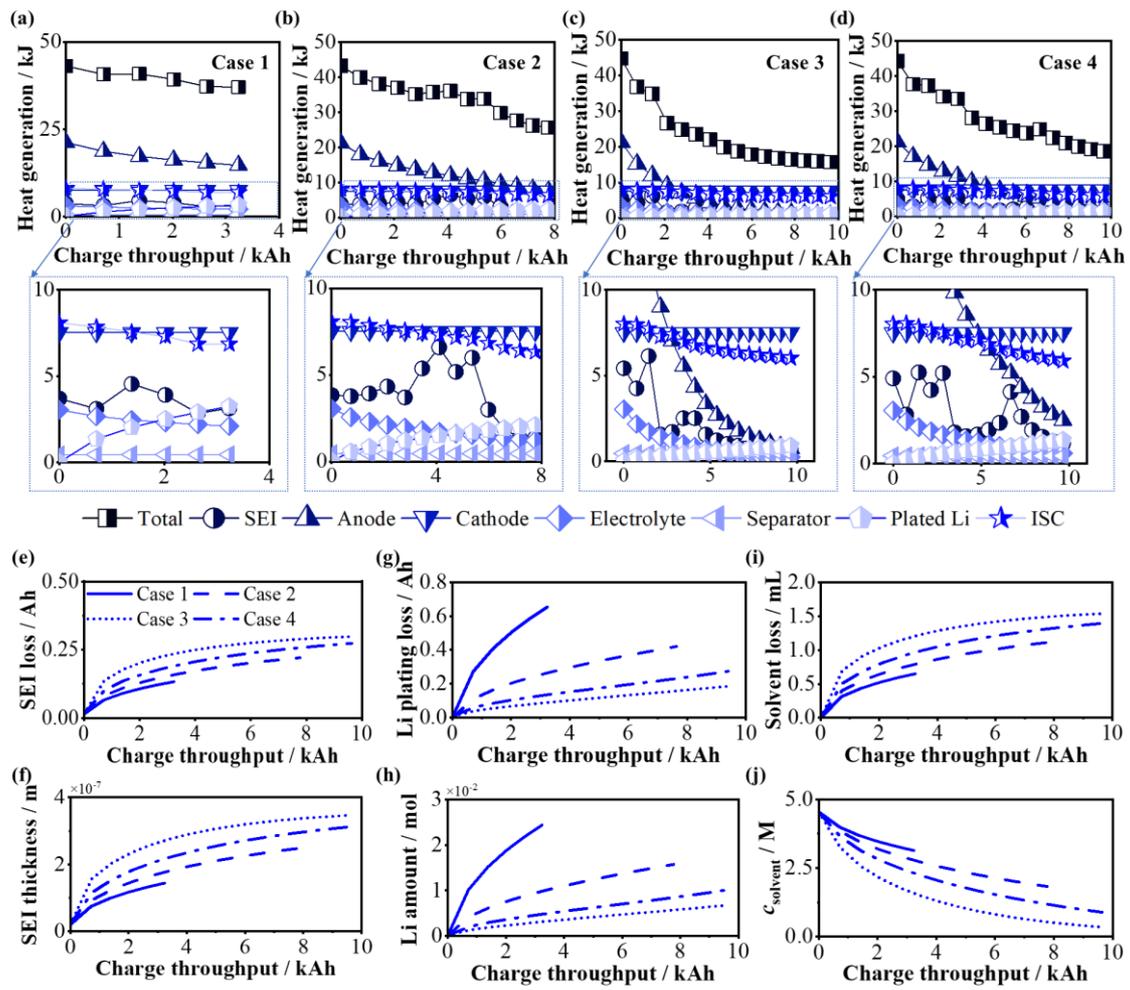

Fig. 6

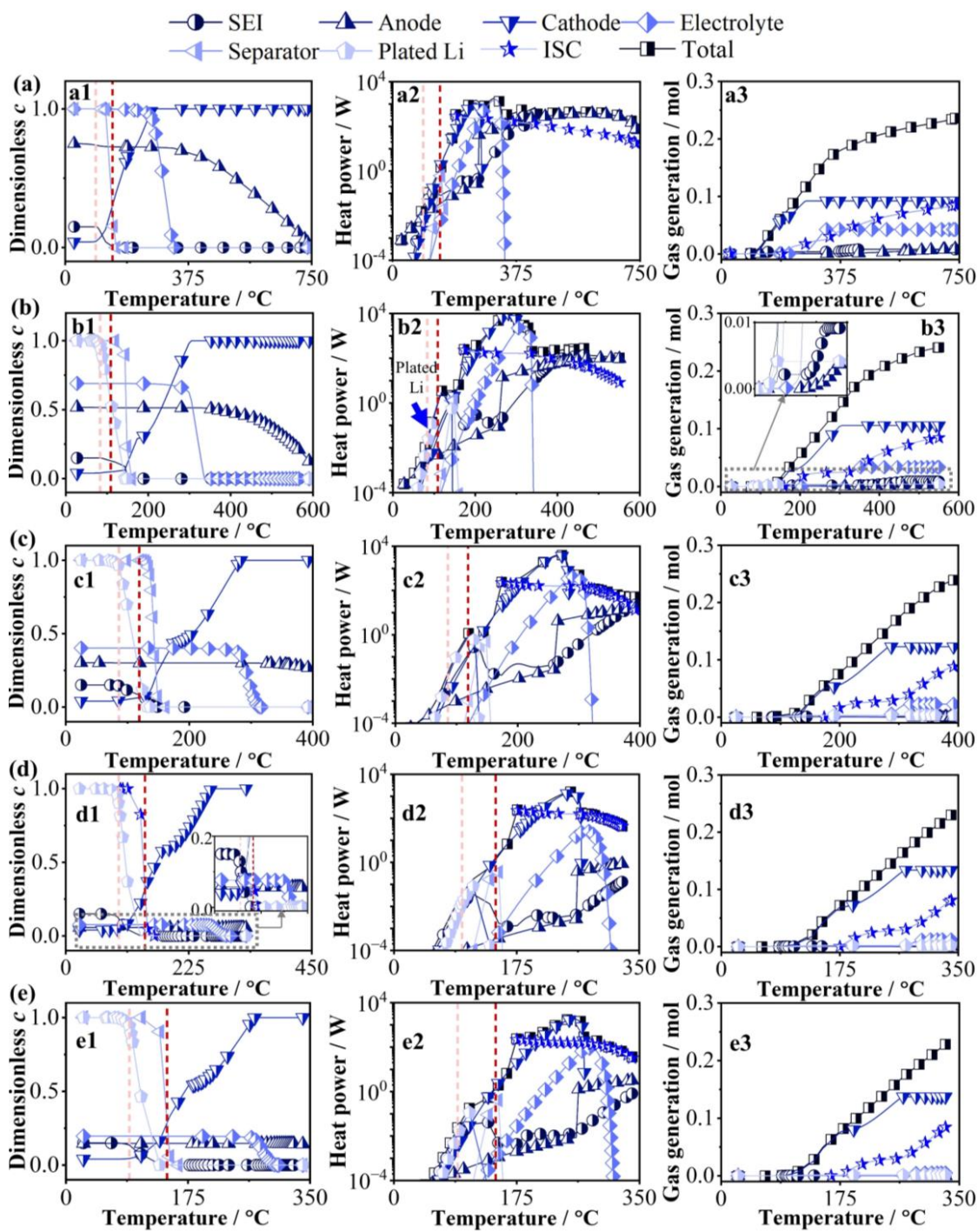

Fig. 7

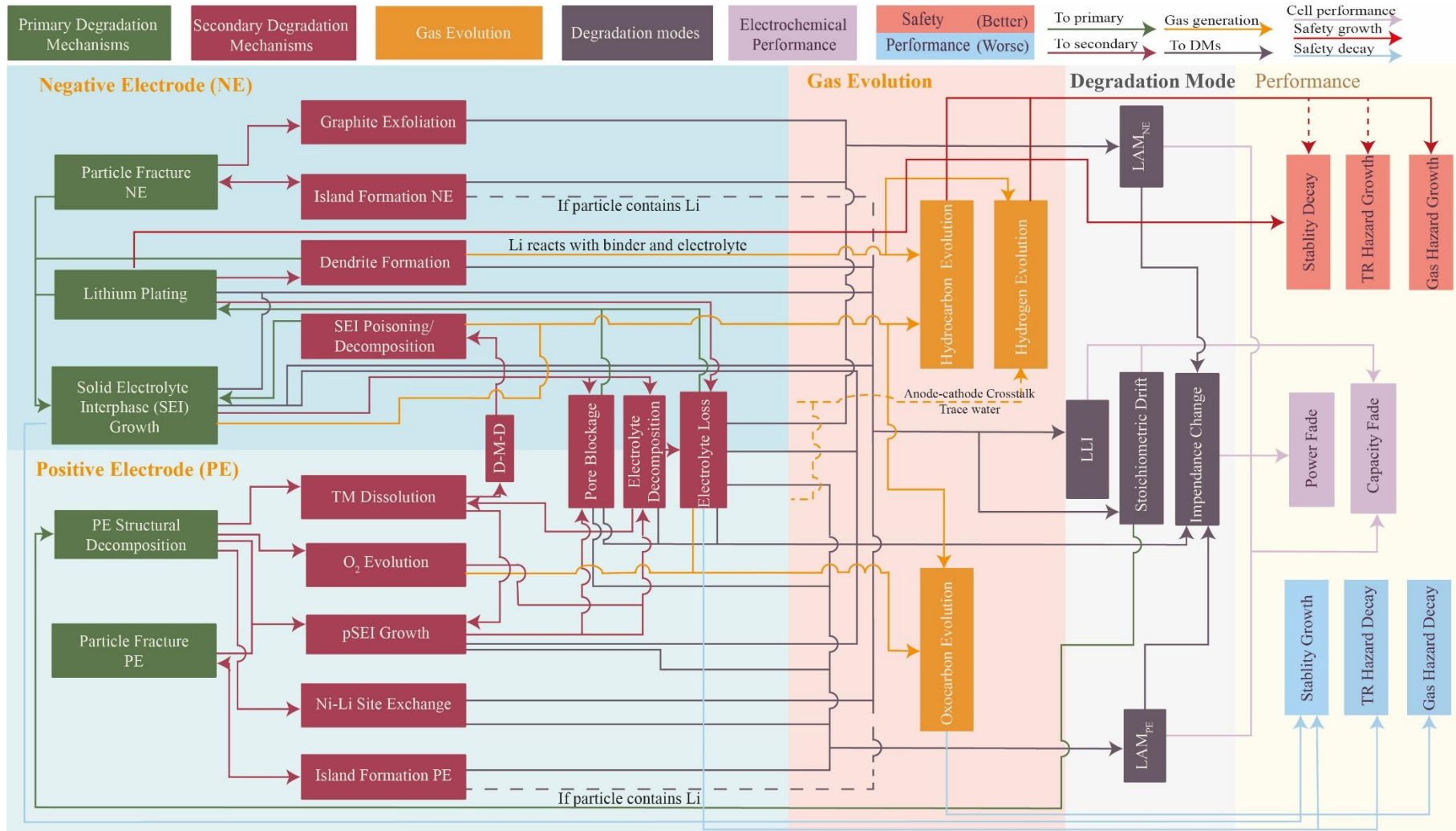

**Supporting information for**

**Mapping safety transitions as batteries degrade: A model-based analysis towards full-lifespan battery safety management**

Xinlei Gao, Ruihe Li, Gregory J. Offer, Huizhi Wang[*]

[1] Department of Mechanical Engineering, Imperial College London, London SW7 2AZ, United Kingdom
[2] The Faraday Institution, Harwell Science and Innovation Campus, Didcot OX11 0RA, United Kingdom

[*]Corresponding author. E-mail: huizhi.wang@imperial.ac.uk

Table S1 Cell specifications and design parameters.

| Cell specifications[#] | |
|---|---|
| Battery type | Cylindrical |
| Operation Voltage (V) | 2.5 - 4.2 |
| Capacity (Ah) | 5 |
| Standard charge rate | 0.3C (1455 mA) |
| Maximum charge rate | 0.3C (0-25 °C) / 0.7C (25-45 °C) |
| Energy density (Wh kg$^{-1}$) | 267 |
| Volumetric energy density (Wh L$^{-1}$) | 752 |
| Mass (g) | 68.3 |
| Cathode active material | Li$_x$Ni$_{0.8}$Mn$_{0.1}$Co$_{0.1}$O$_2$ |
| Anode active material | Li$_x$C$_6$ + SiO$_x$ |
| Electrolyte | EC:EMC 3:7 wt% in LiPF$_6$ |
| Separator | Ceramic coated polyolefin |
| Jellyroll width / height (m) | 2.03·10$^{-2}$ / 6.58·10$^{-2}$ |
| Jellyroll volume (m$^3$) | 2.13·10$^{-5}$ |
| Jellyroll surface area (m$^2$) | 4.84·10$^{-3}$ |
| Cell diameter / height (m) | 2.1·10$^{-2}$ / 71·10$^{-2}$ |
| Casing thickness (m) | 3.4·10$^{-4}$ |
| Cell volume (m$^3$) | 2.43·10$^{-5}$ |
| Negative tab width/ length / thickness (m) | 4.00·10$^{-3}$ / 5·10$^{-2}$ / 1·10$^{-4}$ |
| Positive tab width/ length / thickness (m) | 3.5·10$^{-3}$ / 7·10$^{-2}$ / 1·10$^{-4}$ |
| **Gravimetric and volumetric contributions of components** [1] | |
| Negative electrode coating | Mass: 16.51 g (24.2%) |
| | Volume: 9.49 cm$^3$ (41.2%) |
| Copper foil | Mass: 5.85 g (8.6%) |
| | Volume: 0.66 cm$^3$ (2.9%) |
| Positive electrode coating | Mass: 26.09 g (38.3%) |
| | Volume: 7.98 cm$^3$ (35.1%) |
| Aluminum foil | Mass: 2.87 g (4.2%) |
| | Volume: 1.06 cm$^3$ (4.7%) |
| Separator | Mass: 1.96 g (2.9%) |
| | Volume: 2.07 cm$^3$ (9.1%) |
| Electrolyte (incl. LiPF$_6$) | Mass: 3.80 g (5.6%) |
| | Volume: 3.12 cm$^3$ [##] |
| Casing | Mass: 10.64 g (15.6%) |
| | Volume: 1.35 cm$^3$ (5.9%) |
| Others | Mass: 0.39 g (0.6%) |
| | Volume: 0.1 cm$^3$ (0.4%) |

[#] Value taken from LG M50T datasheet
[##] Electrolyte removed from the calculation as it is soaked into the jellyroll

Table S2 Parameters used for the Heat-Wait-Seek (HWS) mode of EV-ARC testing.

| **Test parameters** | **Values** |
|---|---|
| Starting temperature | 30 °C |
| End temperature | 300 °C |
| Slope sensitivity | 0.02 °C min$^{-1}$ |
| Heat step temperature | 5 °C |
| Wait time | 60 min |
| Data step temperature | 0.1 °C |
| Data step time | 1 s |
| Exo step temperature | 0.1 °C |
| Calculation step temperature | 0.2 °C (10 min seek time) |
| Heater power fraction | 30% |

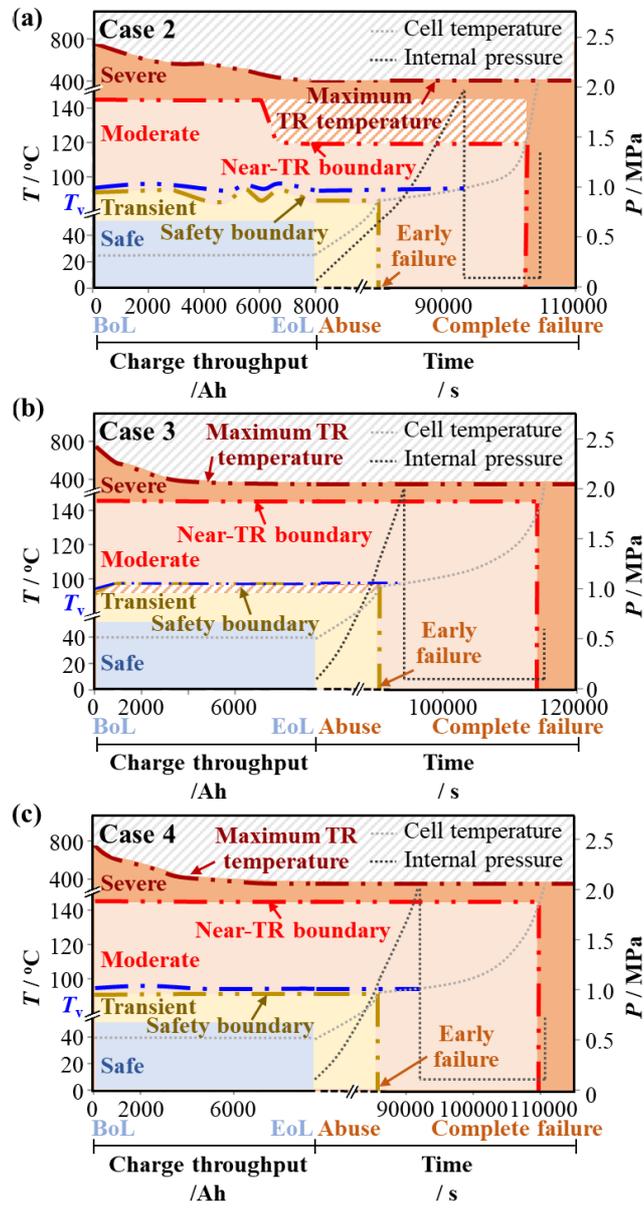

Fig. S1 Dynamic evolutions of safety zones and their boundaries in (a) Case 2, (b) Case 3 and (c) Case 4.

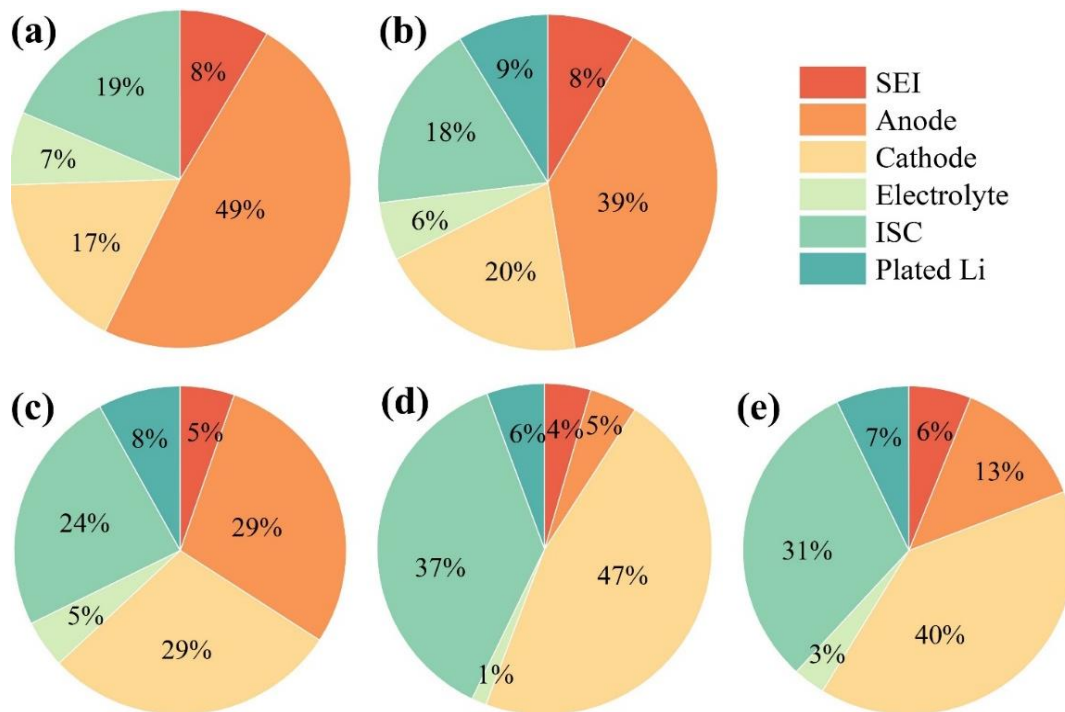

Fig. S2 Percentage contribution of each reaction to the total TR heat generation in the (a) fresh cell and EoL cell in (b) Case 1, (c) Case 2, (d) Case 3 and (e) Case 4.

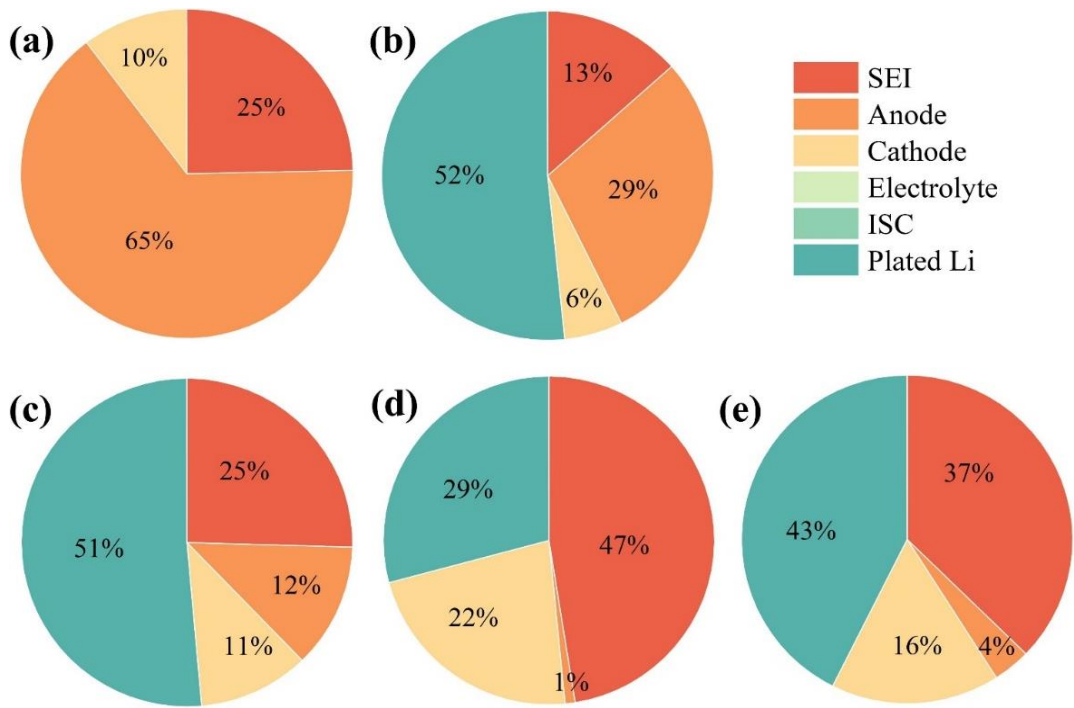

Fig. S3 Percentage contribution of each reaction to the total heat generation at the safety boundary in the (a) fresh cell and EoL cell in (b) Case 1, (c) Case 2, (d) Case 3 and (e) Case 4.

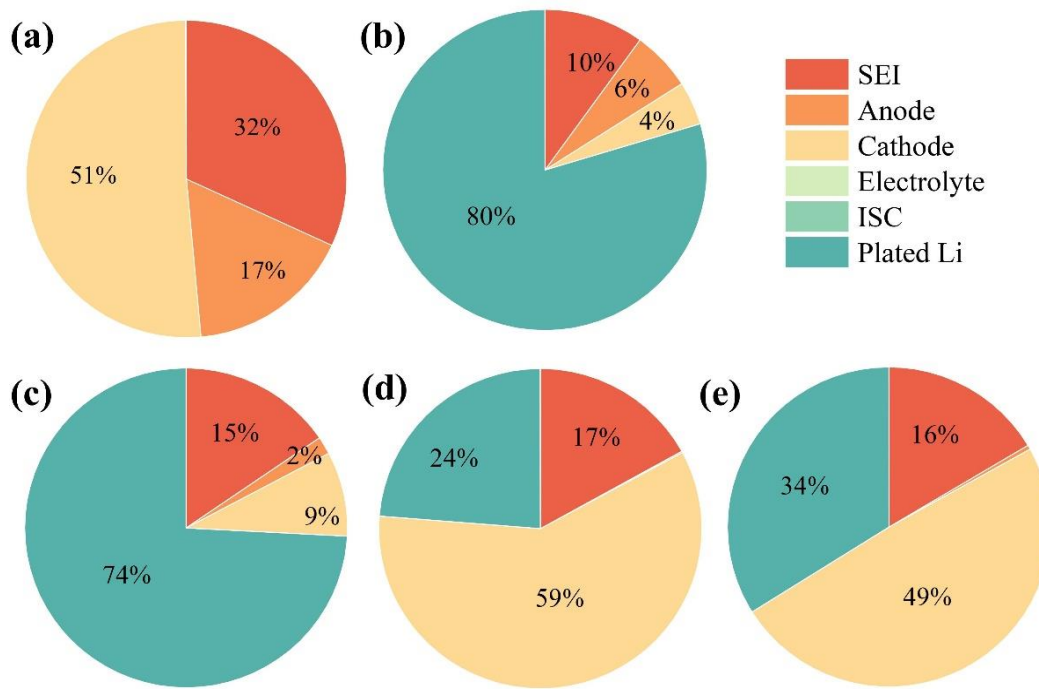

Fig. S4 Percentage contribution of each reaction to the total heat generation at the near-TR boundary in the (a) fresh cell and EoL cell in (b) Case 1, (c) Case 2, (d) Case 3 and (e) Case 4.

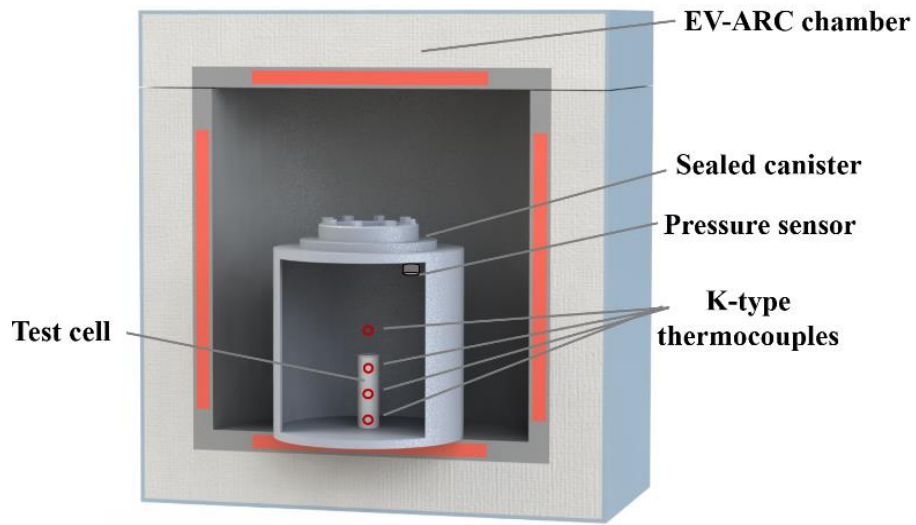

Fig. S5 Experimental set-up for EV-ARC testing.

# Nomenclature

| | | |
|---|---|---|
| $A$ | Area, m$^2$ | |
| $a$ | Surface area to volume ratio, m$^{-1}$ | |
| $B$ | Pre-exponential factor, s$^{-1}$ | |
| $b$ | Stress intensity factor correction | |
| $Cp$ | Specific heat capacity, J kg$^{-1}$ K$^{-1}$ | |
| $c$ | Concentration, mol m$^{-3}$ | |
| $\langle c \rangle$ | Reference concentration, mol m$^{-3}$ | |
| $D$ | Lithium ion diffusivity, m$^2$ s$^{-1}$ | |
| $E$ | Activation energy, J mol$^{-1}$ | |
| $F$ | Faraday constant, C mol$^{-1}$ | |
| $\Delta H$ | Specific enthalpy change, J kg$^{-1}$ | |
| $h$ | heat transfer coefficient, W m$^{-2}$ K$^{-1}$ | |
| $I$ | Current, A | |
| $J$ | Current density, A m$^{-2}$ | |
| $K$ | Modification coefficient | |
| $k$ | Reaction kinetic rate constant, m s$^{-1}$ | |
| $L$ | Thickness, m | |
| $l$ | Length, m | |
| $M$ | Molar mass, kg mol$^{-1}$ | |
| $m$ | Mass, kg | |
| $n$ | Amount of substance, mol | |
| $P$ | Pressure, Pa | |
| $Q$ | Heat generation, J | |
| $\dot{Q}$ | Heat generation rate, W | |
| $q$ | The first reaction function order | |
| $Ratio$ | Dry-out ratio | |
| $R$ | Gas constant, J mol$^{-1}$ K$^{-1}$ | |
| $s$ | The second reaction function order | |
| $T$ | Temperature, K | |
| $t$ | Transference number | |
| $U$ | Voltage, V | |
| $V$ | Volume, m$^{-3}$ | |
| $\bar{v}$ | Partial molar volume, m$^3 \cdot$mol$^{-1}$ | |
| $w$ | Width, m | |
| $Y$ | Mass fraction | |

## *Greek letters*

| | |
|---|---|
| $\alpha$ | Dimensionless indicator for cathode decomposition |
| $\gamma$ | Dead lithium decay constant, s$^{-1}$ |
| $\varepsilon$ | Volume fraction |
| $\epsilon$ | Bruggeman exponent |
| $\eta$ | Overpotential, V |
| $\theta$ | Heat capacity ratio of venting gases |
| $\kappa$ | Ionic conductivity, S m$^{-1}$ |
| $\lambda$ | Exponential term |
| $\nu$ | Poisson's ratio |
| $\xi$ | Empirical factor for ejecta flow |
| $\rho$ | Resistivity, $\Omega \cdot$m |
| $\varrho$ | Density, kg m$^{-3}$ |
| $\sigma$ | Stress, Pa |
| $\tau$ | Time, s |
| $\phi$   | Potential, V                            |
| $\chi$   | Thermodynamic factor                    |
| $\psi$   | Young's modulus, Pa                     |
| $\Omega$ | Partial molar volume                    |
| $\omega$ | Conversion efficiency from electrical energy to heat |

*Superscripts*

| | |
| --- | --- |
| int | (De)Intercalation |
| OCV | Open circuit voltage |
| ref | Reference |
| reg | SEI Regeneration |
| tot | Total |
| * | Dimensionless indicator |
| + | Cation |
| - | Anion |

*Subscripts*

| | |
| --- | --- |
| Aged | Aged cell |
| amb | Ambient |
| anode | Anode intercalated Li reacting with electrolyte |
| avg | Average |
| ARC | Accelerating rate calorimetry |
| cat | Cathode decomposition reaction |
| cell | Battery cell |
| crit | Critical |
| cr | Particle cracking |
| cot | Contact resistance |
| ee | Electrical energy |
| elec | Electrolyte decomposition |
| delith | Delithiation |
| diss | Dissipation |
| dry-out | Electrolyte dry-out |
| evap | Evaporation |
| e | Liquid phase |
| fresh | Fresh cell |
| g | Generated gases |
| ISC | Internal short circuit |
| LAM | Loss of active material |
| lith | Lithiation |
| Li | Li plating |
| loss | Capacity loss |
| max | Maximum |
| n | Anode (positive electrode) |
| out | Outer area |
| Li | Reaction of plated Li with electrolyte |
| p | Cathode (negative electrode) |
| SEI | SEI growth reaction |
| surf | Particle surface |
| sol | Solvent |
| s | Solid phase |
| t | Tangential |
| vent | Venting orifice |
| void | Void space inside the battery |
| 0 | Initial condition |

# Supplementary Method

## DFN model

**Mass conservation for solid phase**

$$\frac{\partial c_{s,k}}{\partial t} = \frac{1}{r^2}\frac{\partial}{\partial r}\left(r^2 D_{s,k}\frac{\partial c_{s,k}}{\partial r}\right) \quad (S1)$$

**Boundary conditions**

$$\left.\frac{\partial c_{s,k}}{\partial r}\right|_{r=0} = 0 \quad -D_{s,k}\left.\frac{\partial c_{s,k}}{\partial r}\right|_{r=R_k} = \frac{J_k^{tot}}{a_k F} \quad (S2)$$

**Charge conservation for solid phase**

$$\frac{\partial}{\partial x}\left(\sigma_{s,k}\frac{\partial \phi_{s,k}}{\partial x}\right) = J_k^{tot} \quad (S3)$$

$$-\sigma_{s,n}\left.\frac{\partial \phi_{s,n}}{\partial x}\right|_{x=0} = -\sigma_{s,p}\left.\frac{\partial \phi_{s,p}}{\partial x}\right|_{x=L} = i_{app}$$

$$-\sigma_{s,n}\left.\frac{\partial \phi_{s,n}}{\partial x}\right|_{x=L_n} = -\sigma_{s,p}\left.\frac{\partial \phi_{s,p}}{\partial x}\right|_{x=L-L_p} = 0 \quad (S4)$$

**Mass conservation for liquid phase**

$$\frac{\partial \varepsilon_k c_{e,k}}{\partial t} = \begin{cases} \frac{\partial}{\partial x}\left(\varepsilon_k^b D_e \frac{\partial c_{e,k}}{\partial x}\right) + (1-t^+)\frac{J_k^{tot}}{F}, k = n,p \\ \frac{\partial}{\partial x}\left(\varepsilon_k^b D_e \frac{\partial c_{e,k}}{\partial x}\right), k = s \end{cases} \quad (S5)$$

$$\left.\frac{\partial c_{e,n}}{\partial x}\right|_{x=0} = \left.\frac{\partial c_{e,p}}{\partial x}\right|_{x=L} = 0 \quad (S6)$$

**Charge conservation for liquid phase**

$$\frac{\partial}{\partial x}\left(\varepsilon_k^b \sigma_e\left(\frac{\partial \phi_{e,k}}{\partial x} - \frac{2(1-t^+)RT}{F}\frac{\partial \log c_{e,k}}{\partial x}\right)\right) = -J_k^{tot} \quad (S7)$$

$$\left.\frac{\partial \phi_{e,n}}{\partial x}\right|_{x=0} = \left.\frac{\partial \phi_{e,p}}{\partial x}\right|_{x=L} = 0 \quad (S8)$$

**Total current density**

$$J_k^{tot} = \begin{cases} J_k^{int} + \sum J_{sr}, k \in \{n,p\} \\ 0, \quad k = s \end{cases} \quad (S9)$$

**Intercalation current density**

$$J_k^{int} = 2Fa_k k_k(T)\sqrt{c_{e,k}c_{s,k,surf}(c_{s,k}^{max} - c_{s,k,surf})}\sinh\left(\frac{F\eta}{2RT}\right) \quad (S10)$$

**Overpotential**

$$\eta_k^{int} = \begin{cases} \phi_{s,n} - \phi_{e,n} - U_n\left(c_{s,n}|_{r=R_n}\right) - \eta_{SEI}, k = n \\ \phi_{s,p} - \phi_{e,p} - U_p\left(c_{s,p}|_{r=R_p}\right), k = p \end{cases} \quad (S11)$$

**Ohmic voltage drop**

$$\eta_{SEI} = \frac{J_n^{tot} L_{SEI}}{a_n \sigma_{SEI}} \quad (S12)$$

**Terminal voltage**

$$V = \phi_{s,P}|_{x=L} - \phi_{s,n}|_{x=0} \quad (S13)$$

**Porosity change**

$$\frac{d\varepsilon_n}{dt} = -a_n \cdot \frac{dL_{tot}}{dt} \quad (S14) \quad L_{tot} = L_{SEI} + c_{Li}\frac{\bar{v}_{Li}}{a_n} + c_{dl}\frac{\bar{v}_{Li}}{a_n} \quad (S15)$$

## Degradation sub-models

**SEI growth**

$$J_{SEI} = Fa_n \frac{c_{sol} D_{sol}(T)}{L_{SEI}} \quad (S16)$$

$$\frac{\partial L_{SEI}}{\partial t} = \frac{c_{sol,0} D_{sol}(T)\bar{v}_{SEI}}{2L_{SEI}} \quad (S17)$$

$$D_{sol}(T) = D_{sol}(T_{ref})\exp\left(-\frac{E_{sol}}{RT} + \frac{E_{sol}}{RT_{ref}}\right) \quad (S18)$$

**Lithium plating**

$$J_{Li} = Fa_n k_{Li}\left(c_{Li}\exp\left(\frac{F\alpha_{a,Li}\eta_{Li}}{RT}\right) - c_e\exp\left(-\frac{F\alpha_{c,Li}\eta_{Li}}{RT}\right)\right) \quad (S19)$$

$$\frac{\partial c_{Li}}{\partial t} = \frac{-j_{Li}}{F} - \frac{\partial c_{dl}}{\partial t} \quad (S20) \quad \frac{\partial c_{dl}}{\partial t} = \gamma_{Li} c_{Li} \quad (S21)$$

$$\gamma_{Li} = \gamma_{Li,0}\frac{L_{SEI,0}}{L_{SEI}} \quad (S22) \quad \eta_{Li} = \phi_s - \phi_e - \eta_{SEI} \quad (S23)$$

**Electrolyte dry-out**

$$\Delta n_{SEI} = \frac{A_{cell}(t)}{\bar{v}_{SEI}}\int_0^{l_n} a_n[L_{SEI}(t+\Delta t) - L_{SEI}]dx \quad (S24)$$

$$Ratio_{dry} = \frac{V_e(t) + \Delta n_{SEI}(\bar{v}_{SEI} - 2\bar{v}_{EC})}{V_e(t)} \quad (S25)$$

$$V_e(t) = A_{cell}(t)[L_n\varepsilon_n(t) + L_s\varepsilon_s + L_p\varepsilon_p] \quad (S26)$$

$$A_{cell}(t+\Delta t) = Ratio_{dry} A_{cell}(t) \quad (S27)$$

**Particle cracking**

$$\sigma_r = \frac{2\Omega\psi}{(1-\nu)}[c_{avg}(R_i) - c_{avg}(r)] \quad (S28)$$

$$\sigma_t = \frac{\Omega\psi}{(1-\nu)}[2c_{avg}(R_i) + c_{avg}(r) - \bar{c}/3] \quad (S29)$$

$$\frac{dl_{cr}}{dN} = \frac{k_{cr}}{t_0}\left(\sigma_t b_{cr}\sqrt{\pi l_{cr}}\right)^{\lambda_{cr}}, \text{when } \sigma_t > 0 \quad (S30)$$

$$\frac{\partial \varepsilon_{s,k}}{\partial t} = \frac{\beta}{t_0}\left(\frac{\sigma_h}{\sigma_c}\right)^{\lambda_{LAM}}, \text{when } \sigma_h > 0 \quad (S31) \quad \sigma_h = (\sigma_r + 2\sigma_t)/3 \quad (S32)$$

## Failure sub-models

**Energy conservation**

$$m_{cell} Cp_{cell}\frac{dT_{cell}}{dt} = \sum \dot{Q}_i + \dot{Q}_{diss} + \dot{Q}_{short} + \dot{Q}_{evap} + \dot{m}_{vent}\Delta H_{vent} \quad (S33)$$

$$\dot{Q}_i(t) = m_i \Delta H_i \frac{dc_i^*(t)}{dt} \quad (S34)$$

$$c_i^*(t) = c_{i,0}^* - \int_0^t \frac{dc_i^*(\tau)}{d\tau}d\tau \quad (S35)$$

$$\dot{Q}_{ISC}(t) = \frac{1}{\Delta t}\omega\left(\Delta H_{ee} - \int_0^t (Q_{ee}\tau)d\tau\right) \quad (S36)$$

$$\dot{Q}_{evap}(t) = \frac{dm_{elec,vap}}{dt}\Delta h_{elec,vap} \quad (S37)$$

$$\dot{Q}_{diss}(t) = hA(T_{ARC} - T_{cell}) \quad (S38)$$

**Reaction kinetics**

$$\frac{dc_i^*(t)}{dt} = B_i \exp\left(-\frac{E_i}{RT_{cell}(t)}\right)f_i(c_i^*(t)) \quad (S39)$$

$$f_i(c_i^*(t)) = (c_i^*(t))^{q_i} \quad (S40)$$

$$\frac{dc_{SEI}^*(t)}{dt} = \frac{dc_{SEI}^{*,d}(t)}{dt} - \frac{dc_{SEI}^{*,reg}(t)}{dt} \quad (S41)$$

$$\frac{dc_{SEI}^{*,reg}(t)}{dt} = K_{SEI}^{reg}\frac{dc_{anode}^*(t)}{dt} \quad (S42)$$

$$L_{SEI,0,aged} = \frac{L_{SEI,aged}}{L_{SEI,0}}L_{SEI,0,fresh}^* \quad (S43)$$

$$c_{anode,0,aged}^* = \frac{c_{sol,aged}}{c_{sol,fresh}}\cdot c_{anode,0,fresh}^* \quad (S44)$$

$$c_{elec,0,aged}^* = \frac{c_{sol,aged}}{c_{sol,fresh}}\cdot c_{elec,0,fresh}^* \quad (S45)$$

$$f_\alpha(\alpha(t)) = \alpha^{q_{cat}}(1-\alpha)^{s_{cat}} \quad (S46)$$

$$f_{Li}(c_{Li}(t)) = q_{Li}c_{Li}(t)\cdot (-\ln(1-c_{Li}(t)))^{\frac{q_{Li}-1}{q_{Li}}} \quad (S47)$$

$$n_{Li} = \frac{K_{Li}\cdot Q_{loss,LiP}}{F} \quad (S48) \quad \Delta H_{Li} = 1.3307\times 10^5 \times \frac{n_{Li}}{m_{anode}} + 9.6881 \quad (S49)$$

**Internal pressure**

$$P_{tot} = P_g + P_{elec} \quad (S50)$$

$$P_g = \frac{n_g RT_g}{V_{void}} \quad (S51)$$

$$n_g = n_{0,g} + \sum_i \int \frac{dn_{g,i}}{dt}dt \quad (S52)$$

$$\frac{dn_{g,i}}{dt} = \frac{K_{i,g}\Delta n_{g,i}}{c_{i,0}^*}\frac{dc_i^*}{dt} \quad (S53)$$

$$\Delta n_{g,i} = \frac{Q_i}{Q_{tot}} \quad (S54)$$

$$P_{elec} = \exp\left(a - \frac{b}{T+c}\right) \quad (S55)$$

**Dynamic pressure evolution**

$$\frac{dP_{cell}}{dt} = \frac{dP_{elec}}{dt} + \frac{dP_{gas}}{dt} \quad (S56)$$

$$\frac{dP_{elec}}{dt} = \frac{RT_{cell}}{M_{elec}V_{void}}\cdot \frac{dm_{elec}}{dt} + \frac{m_{elec}R}{M_{elec}V_{void}}\cdot \frac{dT_{cell}}{dt} \quad (S57)$$

$$\frac{dP_{gas}}{dt} = \frac{RT_{cell}}{V_{void}}\cdot \frac{dn_{gas}}{dt} + \frac{n_{gas}R}{V_{void}}\cdot \frac{dT_{cell}}{dt} \quad (S58)$$

$$\dot{m}_{ori} = \xi A_{vent}\varrho_{ori}u_{ori} \quad (S59)$$

$$\frac{dm_{elec}}{dt} = -\dot{m}_{elec,v} - \varphi_{elec}\dot{m}_{ori} \quad (S60)$$

$$\frac{dn_{gas}}{dt} = \sum_i \frac{dn_{g,i}}{dt} - \frac{\varphi_{gas}\dot{m}_{ori}}{M_{gas}} \quad (S61)$$

Fig. S6 Governing equations and boundary conditions for the physics-based degradation-safety model.

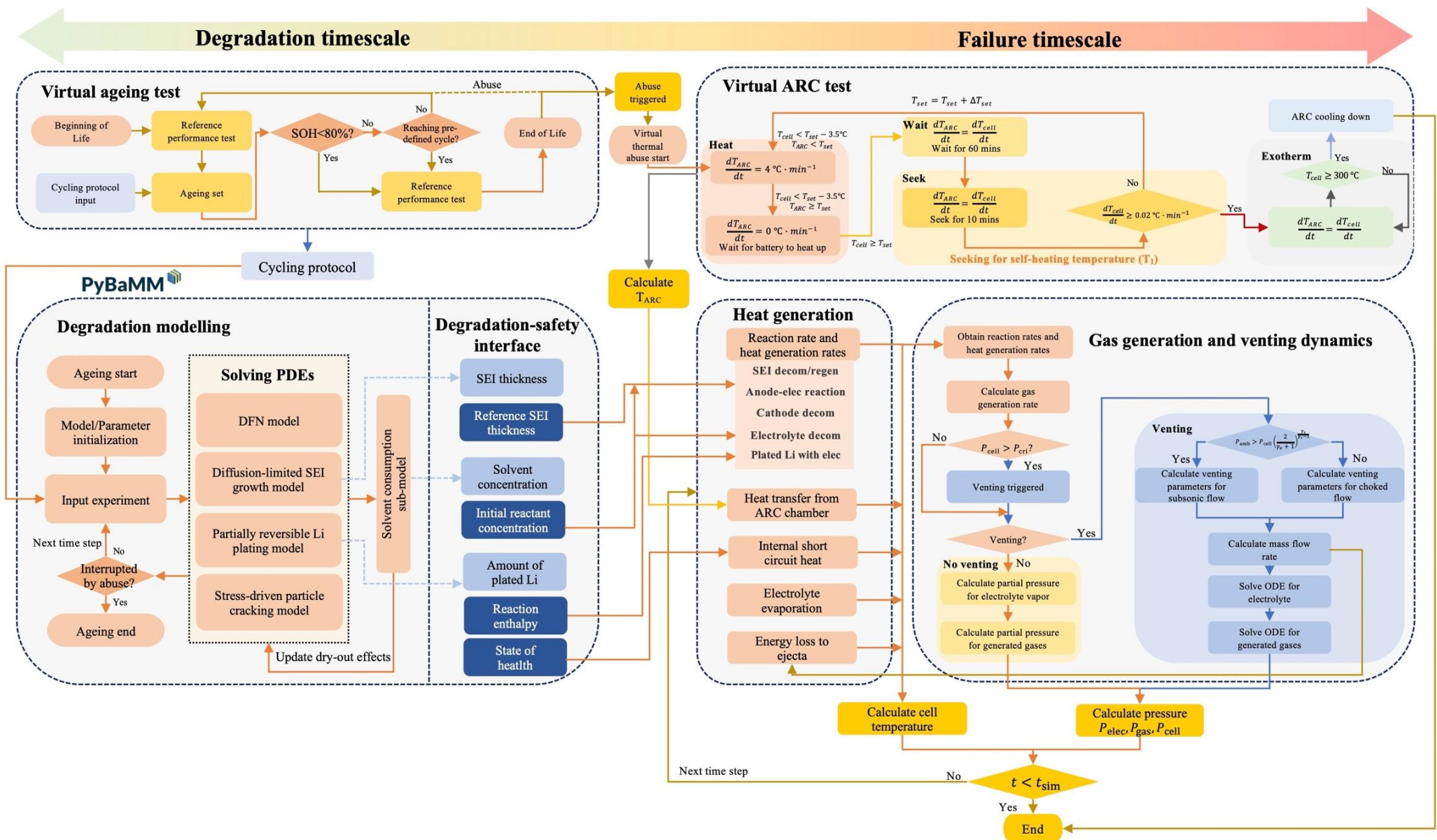

Fig. S7 Flowchart of solving the physics-based degradation-safety model.

Table S3 Input parameters for the degradation-safety model.

| **DFN model** | |
|---|---|
| Current collector thickness / m | $1.6 \cdot 10^{-5}$ (cathode) [1]; $1.2 \cdot 10^{-5}$ (anode) [1] |
| Thickness / m | $7.56 \cdot 10^{-5}$ (cathode) [1]; $1.2 \cdot 10^{-5}$ (separator) [1]; $8.52 \cdot 10^{-5}$ (anode) [1] |
| Electrode length / m | 1.58 [1] |
| Electrode width / m | $6.5 \cdot 10^{-2}$ [1] |
| Mean particle radius / m | $5.22 \cdot 10^{-6}$ (cathode) [1]; $5.86 \cdot 10^{-6}$ (anode) [1] |
| Electrolyte volume fraction | 0.335 (cathode) [1]; 0.47 (separator) [1]; 0.25 (anode) [1] |
| Contact resistance / Ω | 0.01 [1] |
| Electrode solid phase diffusivity / $m^2 \ s^{-1}$ | Eq. (S75) |
| Electrode solid phase conductivity / S $m^{-1}$ | 0.18 (cathode) [1]; 215 (anode) [1] |
| Maximum Li concentration in electrode / mol $m^{-3}$ | 63104 (cathode) [2]; 33133 (anode) [2] |
| Initial Li concentration in electrode / mol $m^{-3}$ | 17038 (cathode) [2]; 29866 (anode) [2] |
| Electrode surface area to volume ratio / $m^{-1}$ | $3.82 \cdot 10^5$ (cathode) [2]; $3.84 \cdot 10^5$ (anode) [2] |
| $Li^+$ diffusivity in the electrolyte / $m^2 \ s^{-1}$ | Eq. (S70) |
| Electrolyte conductivity / S $m^{-1}$ | Eq. (S71) |
| Li transference number | Eq. (S72) |
| Thermodynamic factor | Eq. (S73) |
| Initial (reference) $Li^+$ concentration in electrolyte / mol $m^{-3}$ | 1000 [3] |
| Initial (reference) EC concentration in electrolyte / mol $m^{-3}$ | 4541 [3] |
| Equilibrium potential / V | Eqs. (S62), (S63) for cathode; Eqs. (S64), (S65) for anode |
| Entropy change / V $K^{-1}$ | Eq. (S66) for cathode; Eq. (S67) for anode |
| Exchange current density / A $m^{-2}$ | Eq. (S68) for cathode; Eq. (S69) for anode |
| **Degradation sub-model** | |
| SEI partial molar volume / $m^3 \ mol^{-1}$ | $9.585 \cdot 10^{-5}$ [4] |
| SEI resistivity / Ω m | $2 \cdot 10^5$ [4] |
| Initial SEI thickness / m | $5 \cdot 10^{-9}$ [4] |
| Solvent diffusion activation energy / J $mol^{-1}$ | 37000 [5] |
| Outer SEI solvent diffusivity / $m^2 \ s^{-1}$ | $1.3 \cdot 10^{-20}$ [###] |
| SEI growth activation energy / J $mol^{-1}$ | $3.8 \cdot 10^4$ [2] |
| Initial electrolyte excessive ratio | 1 [3] |
| Molar ratio of lithium to SEI | 2 [3] |
| Dead lithium decay constant / $s^{-1}$ | $1 \cdot 10^{-4}$ [###] |
| Lithium plating kinetic rate constant / m $s^{-1}$ | $5 \cdot 10^{-9}$ [###] |
| Transfer coefficient for Li stripping | 0.35 (anodic) [2]; 0.65 (cathodic) [2] |
| LAM constant proportional term / $s^{-1}$ | $8.33 \cdot 10^{-7}$ ($5.56 \cdot 10^{-6}$ for 40 °C) (anode) [*]; $5.56 \cdot 10^{-6}$ (cathode) [*] |
| Young's modulus / Pa | $3.75 \cdot 10^{11}$ (cathode) [6]; $1.5 \cdot 10^{10}$ (anode) [6] |
| Poisson's ratio | 0.2 (cathode) [6]; 0.3 (anode) [6] |
| Partial molar volume / $m^3 \cdot mol^{-1}$ | $1.25 \cdot 10^{-6}$ (cathode) [6]; $3.1 \cdot 10^{-6}$ (anode) [6] |
| Initial crack length / m | $2 \cdot 10^{-5}$ at both electrodes [6] |
| Initial crack width / m | $1.5 \cdot 10^{-5}$ at both electrodes [6] |
| Cracking rate | $3.9 \cdot 10^{-20}$ at both electrodes [6] |
| Number of cracks per unit area | $3.18 \cdot 10^{15}$ at both electrodes [6] |
| Stress intensity factor correction | 1.12 at both electrodes [6] |
| Paris' law exponential term | 2.2 at both electrodes [6] |
| Critical stress for particle fracture | $3.75 \cdot 10^8$ (cathode) [2]; $6 \cdot 10^7$ (anode) [2] |
| Loss of active material exponential term | 2 at both electrodes [2] |
| **TR failure sub-model** | |
| Cell specific heat capacity / J $kg^{-1} \ K^{-1}$ | 887 [1] |
| Pre-exponential factor / s$^{-1}$ | 1.667·10$^{15}$ (SEI) [7,8]; 0.035 (T < 260 °C), 5 (T >= 260 °C) (anode) [7]; 6.6·10$^{13}$ (cathode) [8]; 3·10$^{15}$ (elec) [7]; 1.3012·10$^{24}$ (pLi) [9]; 1.5·10$^{50}$ (sep) [7] |
| Specific enthalpy / J g$^{-1}$ | 257 (SEI) [7,8]; 1714 (anode) [7]; 300 (cathode); 800 (elec) [7]; $f(n_{Li})$ (pLi) [9]; -233 (sep) [7] |
| Activation energy / J mol$^{-1}$ | 1.4·10$^5$ (SEI) [7,8]; 3.3·10$^4$ (anode) [7]; 1.38·10$^5$ (cathode); 1.7·10$^5$ (elec) [7]; 2.035·10$^5$ (pLi) [9]; 4.2·10$^5$ (sep) [7] |
| Mass of reactant / g | 16.51 (SEI) [1]; 16.51 (anode) [1]; 26.09 (cathode) [1]; 3.8 (elec) [1]; 16.51 (pLi) [1]; 1.96 (sep) [1] |
| Initial dimensionless concentrations for fresh cells | 0.15 (SEI) [7,8]; 0.75 (anode) [7]; 0.04 (cathode) [8]; 1 (elec) [7,8]; 1 (pLi) [9]; 1 (sep) [7] |
| Reaction functions | $c^*_{SEI}$ (SEI); $c^*_{anode} \exp(-L^*_{SEI,0,aged}/L^*_{SEI,0,fresh})$ (anode); $f_\alpha(\alpha(t))$ (cathode), See Eq. (S46); $c^*_{elec}$ (elec); $f_{Li}(c_{Li}(t))$ (pLi), See Eq. (S57); $c^*_{sep}$ (elec); |
| **Gas generation and pressure build up sub-model** | |
| Ambient pressure / Pa | 1.01·10$^5$ |
| Critical pressure / Pa | 2·10$^5$ |
| Molar mass of electrolyte / g mol$^{-1}$ | 25.79 [10] |
| Enthalpy for venting ejecta / J g$^{-1}$ | 100 [###] |
| Cell internal void space / m$^{-3}$ | 2.42·10$^{-6}$ [###] |
| Initial amount of gas / mol | 4.33·10$^{-5}$ [11] |
| Empirical parameters for Antoine equation | EC [10]: a = 6.49; b = 1836.57; c = -102.23 EMC [10]: a = 6.4308; b = 1466.437; c = -49.461 |
| Modification coefficient for gas generation | 0.08 (SEI) [###]; 0.08 (anode) [###]; 2.2 (cathode) [###]; 2.5 (elec) [###]; 0.2 (pLi) [###]; 2 (ISC) [###] |
| Empirical factor for ejecta flow | 0.8 [11] |
| Orifice area / m$^{-2}$ | 1.2·10$^{-5}$ [11] |
| Heat capacity ratio of vent gases | 1.4 [11] |

[###] Fitted.

Electrode OCVs as a function of the stoichiometry of the electrode [12]:

$$U^{OCV}_{p,lith} = -0.7983 \cdot x + 4.513 - 0.03269 \cdot \tanh(19.83 \cdot (x - 0.5424)) - 18.23 \cdot \tanh(14.33 \cdot (x - 0.2771)) + 18.05 \cdot \tanh(14.46 \cdot (x - 0.2776)) \quad (S62)$$

$$U^{OCV}_{p,delith} = -0.7836 \cdot x + 4.513 - 0.03432 \cdot \tanh(19.83 \cdot (x - 0.5424)) - 19.35 \cdot \tanh(14.33 \cdot (x - 0.2771)) + 19.17 \cdot \tanh(14.45 \cdot (x - 0.2776)) \quad (S63)$$

$$U_{\text{n,lith}}^{\text{OCV}} = 0.5476 \cdot e^{-422.4 \cdot x} + 0.5705 \cdot e^{-36.89 \cdot x} + 0.1336 - 0.04758 \\ \cdot \tanh(13.88 \cdot (x - 0.2101)) - 0.01761 \cdot \tanh(36.2 \cdot (x - 0.5639)) \\ - 0.0169 \cdot \tanh(11.42 \cdot (x - 1))$$ (S64)

$$U_{\text{n,delith}}^{\text{OCV}} = 1.051 \cdot e^{-26.76 \cdot x} + 0.1916 - 0.05598 \cdot \tanh(35.62 \cdot (x - 0.1356)) \\ - 0.04483 \cdot \tanh(14.64 \cdot (x - 0.2861)) - 0.02097 \\ \cdot \tanh(26.28 \cdot (x - 0.6183)) - 0.02398 \cdot \tanh(38.1 \cdot (x - 1))$$ (S65)

The entropic changes of both electrodes are given in Eqs. (S66) and (S67), taken from O'Regan et al.[1].

$$\frac{\partial U_{\text{p}}}{\partial T} = 0.04006 \cdot e^{-\frac{(x-0.2828)^2}{0.0009855}} - 0.06656 \cdot e^{-\frac{(x-0.8032)^2}{0.02179}}$$ (S66)

$$\frac{\partial U_{\text{n}}}{\partial T} = -0.111 \cdot x + 0.02901 + 0.3562 \cdot e^{-\frac{(x-0.08308)^2}{0.004621}}$$ (S67)

Exchange current densities are given in Eqs. (S68) and (S69) [2]:

$$J_{0,\text{p}}^{\text{int}} = 3.42 \cdot 10^{-6} \cdot e^{-\frac{1.78 \cdot 10^4}{R}\left(\frac{1}{T} - \frac{1}{298.15}\right)} \cdot c_{\text{e}}^{0.5} \cdot c_{\text{s,surf}}^{0.5} \cdot (c_{\text{s,max}} - c_{\text{s,surf}})^{0.5}$$ (S68)

$$J_{0,\text{n}}^{\text{int}} = 2.668 \cdot e^{-\frac{4 \cdot 10^4}{R}\left(\frac{1}{T} - \frac{1}{298.15}\right)} \cdot \left(\frac{c_{\text{e}}}{\langle c_{\text{e}} \rangle}\right)^{0.208} \cdot \left(\frac{c_{\text{s,surf}}}{c_{\text{s,max}}}\right)^{0.792} \cdot \left(1 - \frac{c_{\text{s,surf}}}{c_{\text{s,max}}}\right)^{0.208}$$ (S69)

Electrolyte phase transport properties are based on the EC:EMC 3:7 wt% LiPF$_6$ [13]:

$$D_{\text{e}} = 10^{-10} \cdot 1010 \cdot e^{1.01 \cdot c_{\text{e}}^{\text{cor}}} \cdot e^{-1560/T} \cdot e^{c_{\text{e}}^{\text{cor}} \cdot (-487)/T}$$ (S70)

$$\kappa_{\text{e}} = 0.1 \cdot 0.521 \cdot (1 + (T - 228)) \cdot c_{\text{e}}^{\text{cor}} \cdot \frac{\left(1 - 1.06 \cdot \sqrt{c_{\text{e}}^{\text{cor}}} + 0.8353 \cdot \left(1 - 0.00359 \cdot e^{\frac{1000}{T}}\right) \cdot c_{\text{e}}^{\text{cor}}\right)}{1 + (c_{\text{e}}^{\text{cor}})^4 \cdot \left(0.00148 \cdot e^{\frac{1000}{T}}\right)}$$ (S71)

$$t_+^0 = -12.8 - 0.612 \cdot c_{\text{e}}^{\text{cor}} + 0.0821 \cdot T + 0.904 \cdot (c_{\text{e}}^{\text{cor}})^2 + 0.0318 \cdot c_{\text{e}}^{\text{cor}} \cdot T - 1.27 \cdot 10^{-4} \cdot T^2 + 0.0175 \cdot (c_{\text{e}}^{\text{cor}})^3 - 3.12 \cdot 10^{-3} \cdot (c_{\text{e}}^{\text{cor}})^2 \cdot T - 3.96 \cdot 10^{-5} \cdot c_{\text{e}}^{\text{cor}} \cdot T^2$$ (S72)

$$\chi = 25.7 - 45.1 \cdot c_{\text{e}}^{\text{cor}} - 0.177 \cdot T + 1.94 \cdot (c_{\text{e}}^{\text{cor}})^2 + 0.295 \cdot c_{\text{e}}^{\text{cor}} \cdot T + 3.08 \cdot 10^{-4} \cdot T^2 + 0.259 \cdot (c_{\text{e}}^{\text{cor}})^3 - 9.46 \cdot 10^{-3} \cdot (c_{\text{e}}^{\text{cor}})^2 \cdot T - 4.54 \cdot 10^{-4} \cdot c_{\text{e}}^{\text{cor}} \cdot T^2$$ (S73)

Solid-phase Li diffusivity for the electrodes:

$$D_{\text{s}} = D_{\text{s}}^{\text{ref}} \cdot e^{-\frac{E_{\text{act}}}{R}\left(\frac{1}{T} - \frac{1}{T_{\text{ref}}}\right)}$$ (S74)

where $D_{\text{s}}^{\text{ref}}$ is calculated by Eq. (S75).

$$\chi \log_{10}(D_{\text{s}}^{\text{ref}}/R_{\text{cor}}) = a_0 \cdot x + b_0 + a_1 \cdot e^{-\frac{(x-b_1)^2}{c_1}} + a_2 \cdot e^{-\frac{(x-b_2)^2}{c_2}} + a_3 \cdot e^{-\frac{(x-b_3)^2}{c_3}} + a_4 \cdot e^{-\frac{(x-b_4)^2}{c_4}}$$ (S75)

The values of constants used in calculating transport properties are listed below[13]:

| Constant | Cathode | Anode |
| --- | --- | --- |
| $a_0$ | - | 11.17 |
| $a_1$ | -0.9231 | -1.553 |
| $a_2$ | -0.4066 | -6.136 |
| $a_3$ | -0.993 | -9.725 |
| $a_4$ | - | 1.85 |
| $b_0$ | -13.96 | -15.11 |
| $b_1$ | 0.3216 | 0.2031 |
| $b_2$ | 0.4532 | 0.5375 |
| $b_3$ | 0.8098 | 0.9144 |
| $b_4$ | - | 0.5953 |
| $c_1$ | 0.002534 | 0.0006091 |
| $c_2$ | 0.003926 | 0.06438 |
| $c_3$ | 0.09924 | 0.0578 |
| $c_4$ | - | 0.001356 |
| $R_{\text{cor}}$ | 2.7 | 3.0321 |

**Supplementary Note 1. Model development**

The degradation-safety model solves the equations in Fig. S6 following the process in Fig. S7, covering both degradation and failure timescales. The "virtual ageing test" provides the cycling protocol inputs to the "degradation model". Degradation modelling involves solving the DFN model together with degradation sub-models of different degradation mechanisms. It is noted that the solvent consumption sub-model only becomes active after one ageing set (normally 78 cycles). After each RPT following each ageing set, a "virtual ARC test" is initiated to simulate the testing conditions inside the EV-ARC during thermal abuse tests, providing thermal boundaries for heat generation modelling. In the "virtual ARC test" block in Fig. S6, under HWS mode (a typical ARC testing mode), the battery cell is heated up step by step until it reaches the self-heating temperature. With a temperature step of 5 °C, the ARC initially undergoes the heating stage, increasing the temperature to the set value at a fixed rise rate of 4 °C/min, and it maintains at the set value if the temperature difference between the cell and ARC is below 3.5 °C. Subsequently, the ARC shifts to the wait stage, where the ARC keeps following the temperature rise of the cell for 60 mins. Then, the ARC enters the seek stage and begins to seek for the occurrence of self-heating by monitoring the temperature rise rate of the cell. A rate exceeding 0.02 °C/min within 10 mins indicates self-heating and terminates the seek stage. The ARC switches to exotherm mode to create an adiabatic environment by following the temperature rise rate of the cell without any further heat compensation. If self-heating is not detected within 20 mins, a new heat-wait-seek loop is initiated with a higher set temperature until self-heating occurs. During exotherm mode, the ARC typically operates under a cooling mode once the cell temperature exceeds 300 °C. The TR onset temperature is determined as the temperature when the temperature rise rate exceeds 60 °C/min. Considering that TR in most batteries often occurs after 300 °C and lasts for a short duration, the simulation of HWS mode terminates once the cell reaches the maximum temperature, and ARC cooling is not simulated.

The "heat generation" modelling solves the energy conservation equation (Eq. (S33)) for cell temperatures, considering reaction heats from thermal decomposition reactions with reaction kinetics updated from the degradation model via the "degradation-safety interface". Internal short circuiting is triggered when the temperature rise rate exceeds 20 °C/min, typically caused by separator melting, resulting in massive heat generation within a short duration ($\Delta t$). The resulting heating power from the electrical energy release is expressed in Eq. (S36). Electrolyte vapor evaporation can absorb heat, with its rate related to the evaporation rate as per Eq. (S37). During TR, the reaction between intercalated Li and electrolyte can cause further SEI formation alongside SEI decomposition [7], giving the expressions of the total SEI reaction rate in Eqs. (S41) and (S42). An exponential term is introduced to the reaction rate of intercalated Li-electrolyte reaction to consider the effect of diffusion through the SEI layer. And the dimensionless thickness of SEI layer at EoL is updated as Eq. (S43), where the initial value at BoL is set to 0.033 [14]. Considering solvent consumption effects, the initial dimensionless concentrations for intercalated Li-electrolyte reaction and electrolyte decomposition, $c^*_{\text{anode,0,aged}}$ and $c^*_{\text{elec,0,aged}}$, are updated with the concentration ratio of the remaining solvent to original solvent as shown in Eqs. (S44) and (S45). Instead of normalised concentration, Eq. (S46) uses conversion degree $\alpha$ for cathode decomposition.

Internal pressure comprises partial pressures from generated gases and electrolyte vapor, solved through Eqs. (S50)-(S55). Venting is triggered when internal pressure exceeds the critical pressure, with flow dynamics calculated as an isentropic nozzle flow for idea gas mixtures [11]. Flow type is determined using Eq. (S76), where the flow is considered subsonic if the inequality holds, otherwise, it is choked flow. Subsonic flow is modelled by Eqs. (S77) to (S81), while choked flow is modelled by Eqs. (S82) to (S84).

$$\frac{P_{\text{amb}}}{P_{\text{cell}}} > \left(\frac{2}{\theta+1}\right)^{\frac{\theta}{\theta-1}} \tag{S76}$$

$$P_{\text{vent}} = P_{\text{amb}} \tag{S77}$$

$$T_{\text{vent}} = T_{\text{cell}} \left(\frac{P_{\text{amb}}}{P_1}\right)^{\frac{\theta-1}{\theta}} \tag{S78}$$

$$\varrho_{\text{vent}} = \frac{P_{\text{vent}} M_{\text{gas}}}{R T_{\text{vent}}} \tag{S79}$$

$$u_{\text{vent}} = Ma \sqrt{\theta \frac{P_{\text{vent}}}{\varrho_{\text{vent}}}} \tag{S80}$$

$$Ma = \sqrt{\frac{2}{\theta-1}\left[\left(\frac{P_1}{P_a}\right)^{\frac{\theta-1}{\theta}} - 1\right]} \tag{S81}$$

$$P_{\text{vent}} = P_{\text{cell}} \left(\frac{2}{\theta+1}\right)^{\frac{\theta}{(\theta-1)}} \tag{S82}$$

$$T_{\text{vent}} = T_{\text{cell}} \left(\frac{2}{\theta+1}\right) \tag{S83}$$

$$Ma = 1 \tag{S84}$$

$$u_{\text{out}} = \begin{cases} \xi u_{\text{vent}}, & \text{if subsonic flow} \\ \xi u_{\text{ori}} - \xi \frac{P_{\text{amb}} - P_{\text{vent}}}{\varrho_{\text{vent}} u_{\text{vent}}}, & \text{if choked flow} \end{cases} \tag{S85}$$

Then the vent flow velocity towards outer area is calculated by Eq. (S85). The mass flow rate can cause a rapid drop in internal pressure after venting, and this dynamic process is described by Eqs. (S56)-(S59). The remaining electrolyte and generated gases can be calculated by considering the loss due to venting, as shown in Eqs. (S60) and (S61). Simulations of heat and gas generation along with venting dynamics are iterated to the next time step until reaching the preset simulation time or until interrupted by the onset of TR, which marks the conclusion of the entire simulation process for degradation and failure timescales within a single ageing path. This entire process is repeated for all ageing paths at different SoH.

**Supplementary Note 2. Model-experiment comparison**

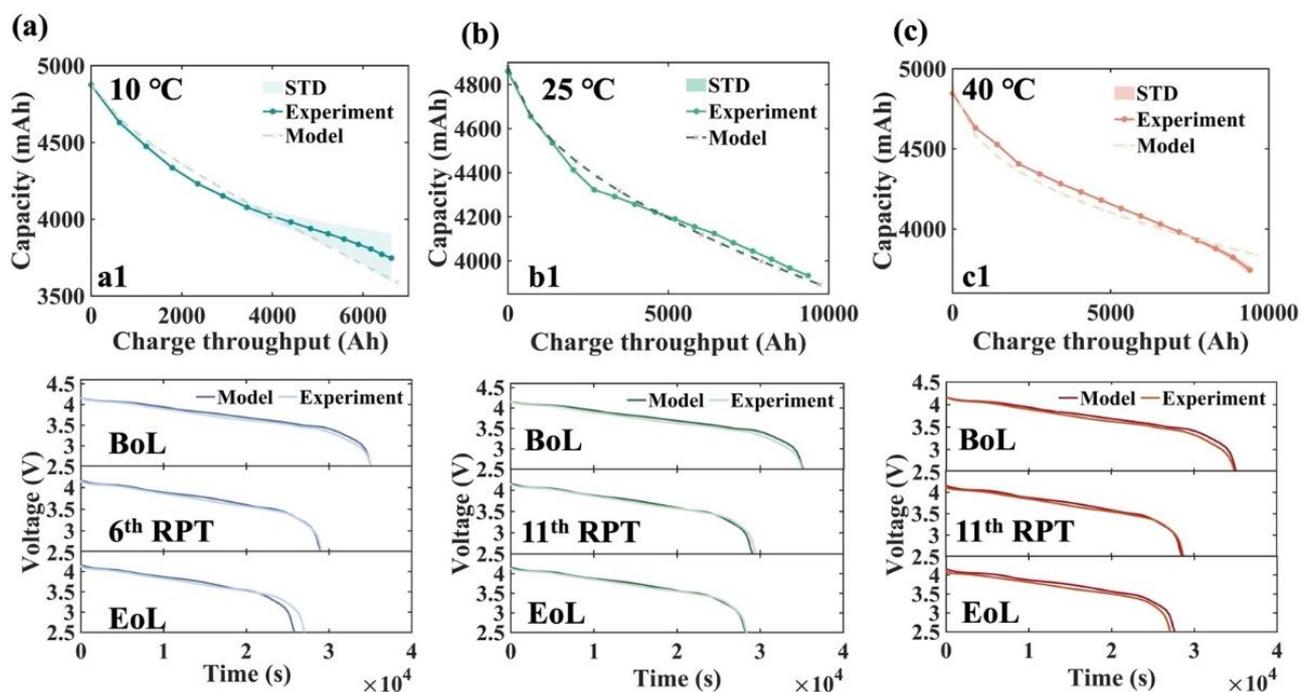

Fig. S8 Model-experiment comparisons for 1/10C capacity retention and voltage profiles in the degradation tests at (a) 10 °C, (b) 25 °C, and (c) 40 °C.

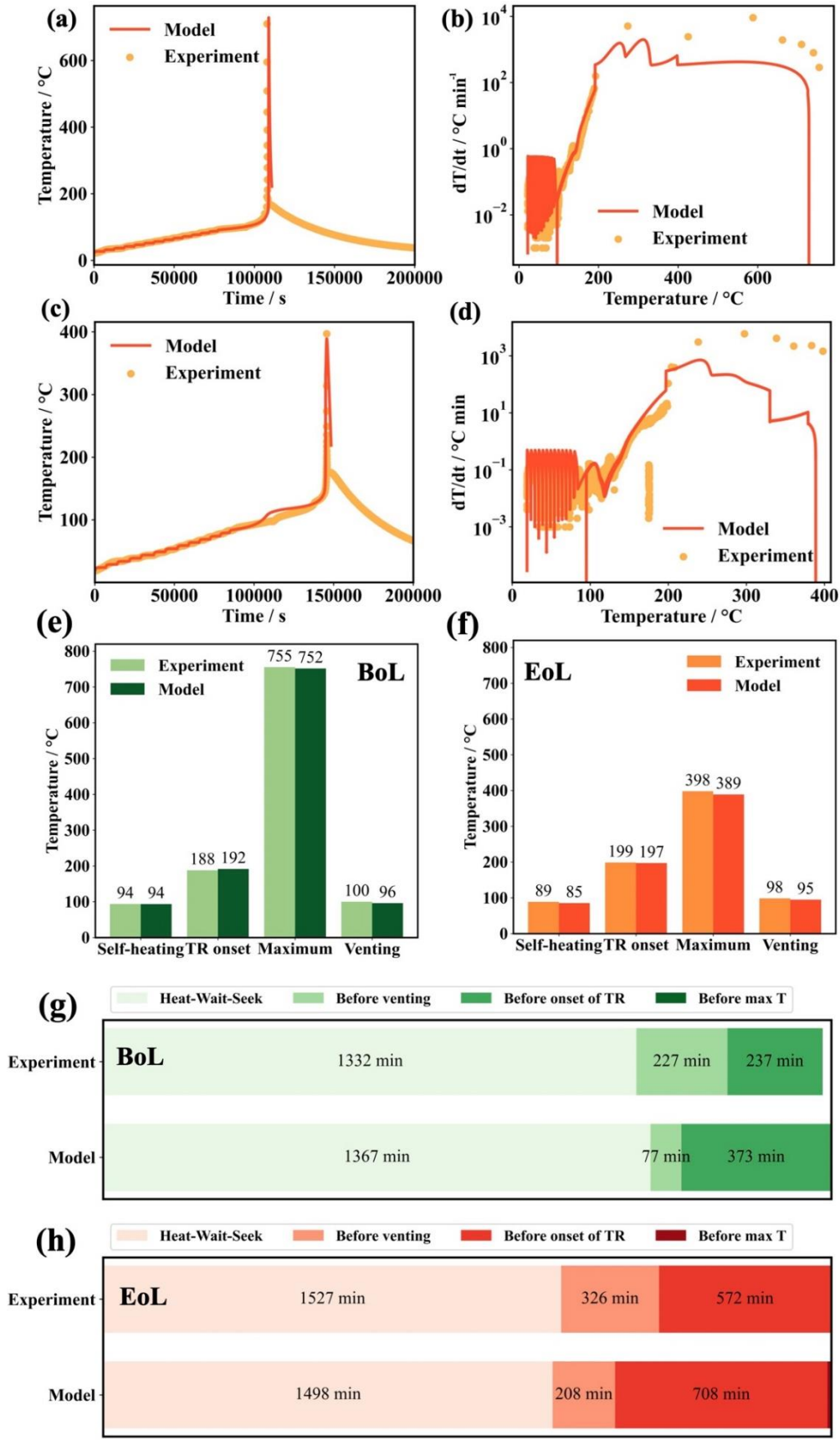

Fig. S9 Model-experiment comparisons for TR failure processes, including (a) cell temperature and (b) temperature rise rate for the fresh cell, (c) cell temperature and (d) temperature rise rate for the EoL cells, characteristic temperatures for the (e) fresh and (f) EoL cells, and TR failure durations for the (g) fresh and (h) EoL cells. The EoL cells were obtained by cycling at 0.3C/1C conditions at 25°C.

**References:**

1. O'Regan, K., Brosa Planella, F., Widanage, W. D. & Kendrick, E. Thermal-electrochemical parameters of a high energy lithium-ion cylindrical battery. *Electrochim Acta* **425**, 140700 (2022).
2. O'Kane, S. E. J. *et al.* Lithium-ion battery degradation: how to model it. *Physical Chemistry Chemical Physics* **24**, 7909–7922 (2022).
3. LI, R., O'Kane, S. E. J., Marinescu, M. & Offer, G. J. Modelling Solvent Consumption from SEI Layer Growth in Lithium-Ion Batteries. *J Electrochem Soc* **169**, 060516 (2022).
4. Safari, M., Morcrette, M., Teyssot, A. & Delacourt, C. Multimodal Physics-Based Aging Model for Life Prediction of Li-Ion Batteries. *J Electrochem Soc* **156**, A145 (2009).
5. Waldmann, T., Wilka, M., Kasper, M., Fleischhammer, M. & Wohlfahrt-Mehrens, M. Temperature dependent ageing mechanisms in Lithium-ion batteries - A Post-Mortem study. *J Power Sources* **262**, 129–135 (2014).
6. Ai, W., Kraft, L., Sturm, J., Jossen, A. & Wu, B. Electrochemical Thermal-Mechanical Modelling of Stress Inhomogeneity in Lithium-Ion Pouch Cells. *J Electrochem Soc* **167**, 013512 (2020).
7. Wang, Y., Ren, D., Feng, X., Wang, L. & Ouyang, M. Thermal kinetics comparison of delithiated Li [ Ni x Co y Mn 1-x-y ] O 2 cathodes. *J Power Sources* **514**, 230582 (2021).
8. Kim, G. H., Pesaran, A. & Spotnitz, R. A three-dimensional thermal abuse model for lithium-ion cells. *J Power Sources* **170**, 476–489 (2007).
9. Ren, D. *et al.* An electrochemical-thermal coupled overcharge-to-thermal-runaway model for lithium ion battery. *J Power Sources* **364**, 328–340 (2017).
10. Fang, Y. J. & Qian, J. M. Isobaric vapor-liquid equilibria of binary mixtures containing the carbonate group -OCOO. *J Chem Eng Data* **50**, 340–343 (2005).
11. Kong, D., Wang, G., Ping, P. & Wen, J. A coupled conjugate heat transfer and CFD model for the thermal runaway evolution and jet fire of 18650 lithium-ion battery under thermal abuse. *eTransportation* **12**, 100157 (2022).
12. Chen, C.-H. *et al.* Development of Experimental Techniques for Parameterization of Multi-scale Lithium-ion Battery Models. *J Electrochem Soc* **167**, 080534 (2020).
13. Landesfeind, J. & Gasteiger, H. A. Temperature and Concentration Dependence of the Ionic Transport Properties of Lithium-Ion Battery Electrolytes. *J Electrochem Soc* **166**, A3079–A3097 (2019).
14. Abada, S. *et al.* Combined experimental and modeling approaches of the thermal runaway of fresh and aged lithium-ion batteries. *J Power Sources* **399**, 264–273 (2018).